\documentclass[prb,twocolumn,amsmath,amssymb,aps]{revtex4-1}
\usepackage{graphicx}
\usepackage{dcolumn}
\usepackage{bm}
\usepackage{textcomp}
\usepackage{float}
\usepackage[colorlinks]{hyperref}
\usepackage{epstopdf}

\newcommand{\FP}{Fabry-P\'erot }
\begin{document}


\title{Contact doping, Klein tunneling, and asymmetry of shot noise in suspended graphene}

\author{Antti Laitinen$^1$, G. S. Paraoanu$^1$, Mika Oksanen$^1$, Monica F. Craciun$^2$, Saverio Russo$^2$, Edouard Sonin$^{3,1}$, and Pertti Hakonen$^1$}

\affiliation{$^1$Low Temperature Laboratory, Department of Applied Physics, Aalto University, 00076 AALTO, Finland}
\affiliation{$^2$Centre for Graphene Science, University of Exeter, EX4 4QL Exeter, United Kingdom}
\affiliation{$^3$Racah Institute of Physics, Hebrew University of Jerusalem, Jerusalem 91904, Israel}

\date{\today}

\begin{abstract}

The inherent asymmetry of the electric transport in graphene is attributed to Klein tunneling across barriers defined by \textit{pn}-interfaces between positively and negatively charged regions. By combining conductance and shot noise experiments we determine the  main characteristics of the tunneling barrier (height and slope) in a high-quality suspended sample with Au/Cr/Au contacts. We observe an  asymmetric  resistance $R_{\textrm{odd}}=100-70$ $\Omega$ across the Dirac point of the suspended graphene at carrier density $|n_{\rm G}|=0.3-4 \cdot 10^{11}$ cm$^{-2}$, while  the Fano factor displays a non-monotonic asymmetry  in the range $F_{\textrm{odd}} \sim 0.03 - 0.1$. Our findings agree with analytical calculations based  on the Dirac equation with a trapezoidal barrier. Comparison between the model and the data yields the barrier height for tunneling, an estimate of the  thickness of the \textit{pn}-interface $d < 20$ nm, and the contact region doping corresponding to a Fermi level offset of $\sim - 18$ meV. The strength of pinning of the Fermi level under the metallic contact is characterized in terms of the contact capacitance $C_c=19 \times 10^{-6}$ F/cm$^2$. Additionally, we show that the gate voltage corresponding to the Dirac point is given by the work function difference between the backgate material and graphene.

\end{abstract}

\maketitle

\section{Introduction} \label{intro}

Klein tunneling is one of the most spectacular effects of relativistic quantum field theory described by the Dirac equation. This tunneling phenomenon, present even in the regime of impenetrable barriers, leads to peculiar transport properties of graphene. Klein tunneling is the backbone of  transport due to evanescent modes causing the observed pseudodiffusive behavior of ballistic graphene samples  \cite{Tworzydio2006,Katsnelson2006b}. The bimodal distribution of transmission eigenvalues in ballistic graphene coincides with a diffusive conductor, which results in shot noise non-distinguishable from diffusive mesoscopic conductors. Furthermore, the evanescent modes lead to a minimum conductivity of $\frac{4e^2}{\pi h}$ in the ballistic regime \cite{katsnelson2006a}. Evidence of these Klein tunneling phenomena have been obtained from observations of charge transport and shot noise in a graphene sheet with ballistic characteristics \cite{Miao2007a, Danneau2008,Du2008}.

The most commonly employed assumption in the analysis of the conductance and shot noise of ballistic graphene has been to consider the carbon layer underneath the electrodes as strongly doped, and this can be modelled using a rectangular electrostatic potential \cite{Katsnelson2006b, Tworzydio2006, Sonin2008}. In reality, this assumption suffers of severe limitations as in a real device the charge density varies continuously and the rate of change is governed by the screening length.  Various theoretical models for finite-slope  potentials  have been analyzed   for $pn$-interfaces in graphene \cite{Cheianov2006, Shytov2008, Cayssol2009}.
All the models have predicted asymmetry in transport properties with respect to the gate voltage,\textit{ i.e.} whether the charge carriers are electrons or holes. In recent experiments, such asymmetry has been observed \cite{Huard2007, Huard2008, Stander2009}. In the ballistic regime, this asymmetry is attributed to the Klein tunneling \cite{Stander2009,rickhaus2013,Oksanen2014} while scattering by charged impurities \cite{Novikov2007, Chen2008a} plays also a role in the diffusive regime. Furthermore, evidence of Klein tunneling has been reported in conductance experiments in confined geometries displaying phase-coherent and double-junction interference effects \cite{Young2009, Young2011}. Sharp $pn$-interfaces have also been achieved in non-suspended samples using air-bridge type gates \cite{LiuG2008,Gorbachev2008}. A full understanding of contact issues is of vital importance for the development of novel electrical components using graphene and other 2-dimensional materials \cite{ferrari2015}. In particular, detailed understanding of $pn$-interfaces is critical for optoelectronics components \cite{tielrooij2015}.

The asymptotic carrier transport in Klein tunneling is bound to be affected by the strong influence of the metal contacts on graphene. A simple contact model was formulated by Giovannetti \textit{et al.} \cite{giovannetti2008} who also performed DFT calculations concerning the involved work functions. In this paper, we generalize this model to include the effects of the applied backgate voltage, and we combine the resulting model with tunneling calculations based on the Dirac equation in order to  obtain  a comprehensive transport model for analyzing electrical conduction in a ballistic, suspended graphene sample. We employ a trapezoidal form for the tunneling barrier which we show how to treat analytically \cite{Sonin2009}. By using conductance and shot noise experiments performed  on a high-quality suspended graphene sample, we can determine the barrier parameters and their relation to the doping of the graphene by a metallic contact.  In comparison with DFT calculations \cite{giovannetti2008}, we find a semiquantitative agreement for the graphene-modified metal work functions as well as for the distance between the charge separation layers which govern the contact capacitance between the metal and graphene.

The experiment and the agreement with the theoretical model confirm the existence of Klein tunneling in graphene. Our method works well even in the situation in which the work functions of the contact metal and graphene differ by a relatively small amount (tens of meV). As such, our results suggests a novel method to find the work function of materials. Note that the work function difference between two metals is not measurable directly: the standard way for its determination is the use of Kelvin probe force microscopy, where the electrical capacitance between the metal and a probe is varied in order to induce a measurable AC current. Our results demonstrate that there exists a gating effect in the position of the Dirac point of the suspended graphene due to the work function of the backgate.  Thus, by using a material as backgate for graphene and measuring the gate voltage corresponding to the Dirac point one can get a simple DC measurement of the work function.

The paper is organized as follows: after the present introduction as Sect. \ref{intro}, we present a complete theoretical treatment of the problem of gated suspended graphene with metallic contacts in  Sect. \ref{cond}. The structure of our samples are detailed in Sect. \ref{samplenoise} together with the employed methods for shot noise measurements. Our experimental results and their analysis are presented in Sect. \ref{experiment}. The implications of the results are discussed in Sect. \ref{disc} jointly with a comparison to other works.

\section{Theoretical background} \label{cond}

\subsection{Electrochemical model for suspended graphene samples}  \label{parameters}

When a graphene sheet is brought in contact with a metal, electrons will flow between them in order to equilibrate the Fermi level. This effect has been analyzed in detail in Refs. \onlinecite{giovannetti2008} and \onlinecite{Xia2011}.
We proceed beyond these works and introduce a complete electrochemical model for a suspended graphene sample with metallic contact electrodes. Our model takes into account consistently the effect of the backgate on the doping of the graphene under the metal. We show that this effect can be neglected only if the difference between the work function of the  metallic contact electrode and the graphene layer is large enough (\textit{i.e.} in the limit of high barrier, defined below Eq. (\ref{p01})). If this is not the case, the contact-region doping  acquired from the backgate voltage has to be included in the transport calculations. On top of the electrostatic contributions, our model indicates that the work function difference between the backgate and graphene enters the equilibrium charge density, which, in particular, leads to a small shift in the Dirac point of the suspended part of the graphene. Using electrochemical equilibrium conditions, we derive analytical equations that incorporate all these effects.

\begin{figure}[htb]
\centering
\includegraphics[width=70mm]{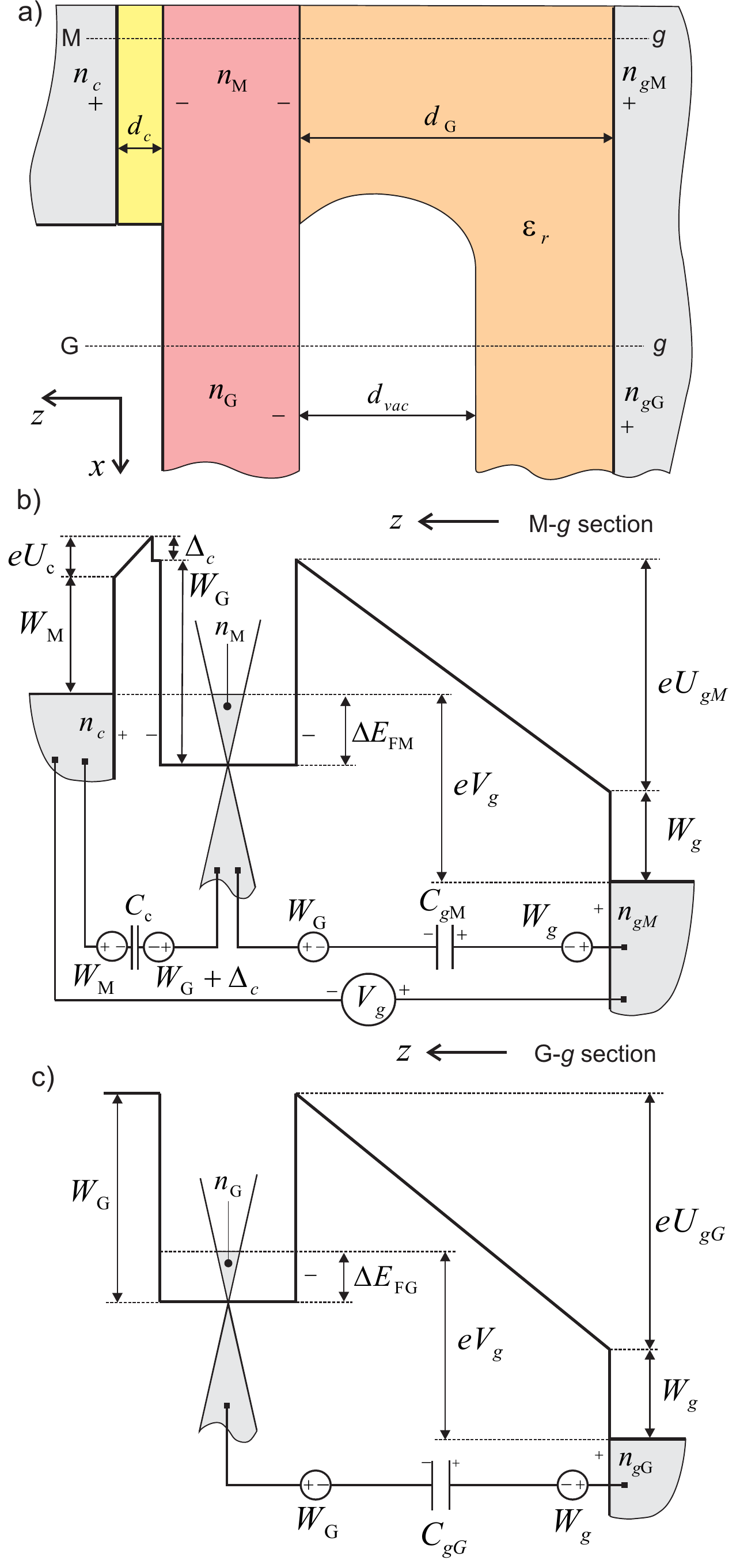}
\caption{ (Color online) Schematic of the sample and of the electrochemical potentials. a)
Section perpendicular to the graphene sheet (pink), showing the metal M (grey) on top, a contact layer (yellow) of width $d_c$, the support dielectric (beige), and the back gate (grey) $g$, at a distance $d_{\rm G}$ under the graphene layer. As seen in the cross section, we assume that the underetching of SiO$_2$ below the graphene can be regarded as small. b) Schematic of the electrochemical potentials for the graphene under the metal (along the line M-$g$). c) Schematic of the electrochemical potential for the suspended graphene (along the line G-$g$). For definition of other symbols, see text.}
\label{fig1}
\end{figure}

In Fig.~\ref{fig1} a) we present a schematic sideview of the sample, showing the region of contact with the metal and the graphene as well as the backgate. For clarity, we have exaggerated the size of some components, so the figure is not to scale. On the sample chip, the graphene sheet is placed at a distance of  $d_{\rm G}  = 300$ nm
from the backgate. The sheet is supported by a SiO$_2$ insulating layer ($\epsilon_r=3.9$), which is partially etched away under the graphene. This results in a vacuum gap of height $d_{\rm vac} = 150$ nm. Further details on the sample and the metallic contacts are found in Sect. \ref{samplenoise}.

In our sample structure, the capacitance per unit area $C_{g{\rm M}}$ between the graphene and the backgate in the support region (graphene under the metal) is estimated from the regular parallel plate formula
\begin{equation}
C_{g{\rm M}} = \frac{\epsilon_{0}\epsilon_{r}}{d_{\rm G}},
\end{equation}
which yields
$C_{g{\rm M}} \sim 1.2 \times 10^{-8}$ F/cm$^2$ using the above mentioned values for $d_{\rm G}$ and $\epsilon_r$.
In the suspended region, the graphene capacitance per unit area against the backgate $C_{\rm gG}$ can be calculated using the formula for two capacitors in series: a vacuum capacitor with a plate separation of $d_{\rm vac}$ and a capacitor with the spacing $d_{\rm G}-d_{\rm vac}$ filled with dielectric material having $\epsilon_r$. This results in the capacitance per unit area
\begin{equation}
C_{\rm gG}=\frac{\epsilon_{0}}{d_{\rm G} +(d_{\rm G}-d_{\rm vac})(\epsilon_r^{-1} - 1)} .
\end{equation}
In our calculations the actual value for the suspended part capacitance is taken from our \FP measurements $C_{\rm gG} = 4.7 \times 10^{-9}$ F/cm$^{2}$  \cite{Oksanen2014}, which agrees well with the above theoretical value.

Next, we analyze the electrochemical potentials that appear in this experimental setting. We take two cuts through Fig.~\ref{fig1} a), one across the dashed line M-$g$, and the other across the dashed line G-$g$, and we represent the spatial variation of the electrochemical potential in Fig.~\ref{fig1} b) and Fig.~\ref{fig1} c), respectively. The work function of the pristine graphene is denoted by $W_{\rm G}$. At the contact with the metal, this work function may be modified by a small shift, see Ref. \onlinecite{giovannetti2008}. The work function of the contact metal on top of the graphene is denoted by $W_{\rm M}$, while the work function of the backgate is $W_{g}$. Note that according to Volta's rule, the contact potential at the end of a circuit is determined only by the work functions of the circuit elements at the end; therefore no other work function, for example corresponding to various other metals along the measurement chain, can enter in this problem.
The difference between the work function of the gate and that of graphene is denoted by $e V_{\rm Dirac} = W_{g}- W_{\rm G}$: this quantity will turn out to be the shift of the Dirac point of the suspended region of graphene due to the backgate work function. Following the common practice in graphene research, gate voltages in the following formulas will be measured with respect to the Dirac point, with the corresponding shifts defined as
\begin{equation}
\delta V_{g} = V_{g} - V_{\rm Dirac}.
\end{equation}

We first solve the problem of finding the surface charge distribution for the electrochemical potentials presented in Fig.~\ref{fig1} b).
At the contact between the metal and the graphene, the electrons will move from the electrode with the lower work function into the electrode with the higher work function. As a result, a surface charge distribution $e n_{c}$ will appear in the contact region, producing an electrostatic potential $U_{c} = e n_{c}/C_{c}$ across the contact capacitance $C_{c}$, the magnitude of which reflects the spatial variation of the surface charge. According to DFT calculations, $C_c \simeq 10^{-5}$ F/cm$^2$ \cite{giovannetti2008}.
A relevant parameter for charge transfer between  the metal and graphene is the difference between the work functions of the graphene under the metal and the work function of the metal, which is described by
\begin{equation}
\chi = W_{\rm G} + \Delta_{c} - W_{\rm M},
\end{equation}

\noindent where $W_{\rm M}$ is the work function of the metallic contact material and $W_{\rm G} + \Delta_{c}$ denotes the modified work function of the graphene under the contact.
Another electrostatic potential $U_{g{\rm M}} = en_{g{\rm M}}/C_{g{\rm M}}$
is established across the capacitance $C_{g{\rm M}}$, with a surface charge $e n_{g{\rm M}}$ on the gate. The total particle-number surface  density $n_{\rm M}$ in the graphene layer in the contact region is therefore
\begin{equation}
n_{\rm M} = n_{g{\rm M}} + n_{c}.\label{fm0}
\end{equation}

The first equilibrium condition is  obtained from the condition that the difference between the Fermi level of the metal and that of the gate equals $e V_{g}$. This condition does not formally involve the characteristic density of states  of graphene. Hence, as shown  by the circuit  schematics below the Fermi level diagram in Fig.~\ref{fig1} b),  it can be regarded as a pure electrostatic condition. It states
\begin{equation}
e \delta V_{g} = \chi + e U_{g{\rm M}} - e U_{c}. \label{fm1}
\end{equation}

 The second equation governing the equilibrium involves the intrinsic properties of graphene, and it can be obtained by using the condition that the Fermi levels of the metal and the graphene under the metal coincide:
\begin{equation}
\Delta E_{\rm FM} + e U_{c} = \chi. \label{fm2}
\end{equation}

In the graphene under the metal, where the linear graphene  bands are supposed to persist, the relation between the number of negatively-charged carriers per unit area $n_M$ and the shift in the Fermi  level $\Delta E_{\rm FM}$ is given by
\begin{equation}
\Delta E_{\rm FM} = \hbar v_{\rm F}{\rm sgn \left[n_{\rm  M}\right]}\sqrt{\pi |n_{\rm  M}|}, \label{fm3}
\end{equation}

\noindent where $v_F = 1.1 \times 10^6$ m/s is the Fermi velocity.
It is useful to introduce a constant $\zeta_{\rm F}$ relating the Fermi speed and the fundamental constants $\hbar$ and $e$:
\begin{equation}
\zeta_{\rm F}= \frac{\sqrt{\pi}\hbar v_{\rm F}}{e}.
\end{equation}

\noindent We propose to call this quantity Fermi electric flux. This constant is related to the concept of quantum capacitance (for graphene, see Ref. \onlinecite{fang2007}) and to the fine structure constant of graphene, as detailed in Appendix \ref{appendix_xiF}. The Fermi electric flux determines the energy shifts produced by graphene as it is inserted in to an electrical circuit. For $v_{\rm F} = 1.1 \times 10^6$ m/s, the equation yields $\zeta_{\rm F} = 1.283 \times 10^{-7}$ V$\cdot$cm.

Combining now Eqs. (\ref{fm0}-\ref{fm3})
above, and choosing the proper physical solution, we obtain the final result for the shift of the energy level of graphene under the metal,
\begin{widetext}
\begin{equation}
\Delta E_{\rm FM} = {\rm sgn}\left[\delta V_{g} + \frac{\chi C_{c}}{eC_{g{\rm M}}}\right]
\left\{-\frac{C_{c}+C_{g{\rm M}}}{2}\zeta_{\rm F}^{2}+\sqrt{\left(\frac{C_{c}+C_{g{\rm M}}}{2}\zeta_{\rm F}^{2}\right)^2
+\zeta_{\rm F}^2 C_{c} \left\vert \chi + \frac{C_{g{\rm M}}}{C_c}
e\delta V_{g}
\right\vert}\right\}. \label{sol1}
\end{equation}
\end{widetext}
An experimentally relevant limit for Eq. (\ref{sol1}) is the case of a very large contact capacitance $C_c$: if the material parameters are such that $(C_{c}+C_{\rm M})\zeta_{\rm F}^2 \gg |\chi|$ and the charge induced by the gate voltage is relatively small, then $eC_{g{\rm M}}|\delta V_{g}|\ll C_{c}|\chi|$, and we obtain
$\Delta E_{\rm FM}\approx \chi$ from Eq. (\ref{sol1}). In this situation, the large contact capacitance locks the position of the Dirac point of the graphene under the metal to a value $\Delta E_{\rm FM}$ set by the parameter $\chi$.

In our experimental setup, in fact, $\chi$ remains relatively small, which makes the quantity $eC_{M}\delta V_{g}/C_{c}$ comparable with $\chi$ at large values of $\delta V_g$. Consequently,
the Fermi level can even reach the Dirac point of the graphene under the metal $\Delta E_{\rm FM}=0$. As the latter condition does require substantial gate voltages, we could not reach this regime in our experiment.

Next we turn to the suspended part of the graphene. In this region, the charge accumulated on the surface of the backgate $n_{g{\rm G}}$, which produces a voltage drop $U_{g{\rm G}}=en_{g{\rm G}}/C_{g{\rm G}}$, is exactly compensated by the charges on the graphene side having the  particle-number surface density $n_{\rm G}$,
\begin{equation}
n_{\rm G} = n_{g{\rm G}}.\label{fg2}
\end{equation}
Then, the condition that the electrochemical potential difference between the Fermi level of graphene and the Fermi level of the backgate is $e V_{g}$ reads
\begin{equation}
e\delta V_{g}=e U_{g{\rm G}} + \Delta E_{\rm FG},\label{fg3}
\end{equation}
where
\begin{equation}
\Delta E_{\rm FG} = \hbar v_{\rm F}{\rm sgn \left[n_{\rm  G}\right]}\sqrt{\pi |n_{\rm  G}|}. \label{fg4}
\end{equation}
in parallel to Eq. (\ref{fm3}). From Eqs. (\ref{fg2}-\ref{fg4}), we find for the Fermi level shift in the suspended part:
\begin{widetext}
\begin{equation}
\Delta E_{\rm FG} = {\rm sgn}\left[\delta V_{g}\right]
\left\{-\frac{C_{\rm gG}}{2}\zeta_{\rm F}^{2}+\sqrt{\left(\frac{C_{\rm Gg}}{2}\zeta_{\rm F}^{2}\right)^2
+ e C_{\rm gG}\zeta_{\rm F}^2\vert \delta V_{g}\vert}\right\}.\label{sol2}
\end{equation}
\end{widetext}

Note that $\Delta E_{\rm FG}=0$ corresponds to $\delta V_{g}=0$,\textit{ i.e.}
$V_{g}=V_{\rm Dirac}$ is indeed the Dirac point of the suspended graphene.
For $\Delta E_{\rm FG}$, typical experimental conditions fulfil
$e\vert \delta V_{g}\vert \gg \vert \chi \vert \gg C_{\rm gG}\zeta_{\rm F}^2$, which allows one to approximate $\Delta E_{\rm FG} \approx {\rm sgn}\left[\delta V_{g}\right]\zeta_{\rm F}\sqrt{eC_{\rm gG}\vert \delta V_{g}\vert}$.
This limit was assumed in the previous analysis by Sonin \cite{Sonin2009}.

The difference of the Fermi energies of the suspended graphene and the graphene under the contact results in an electrostatic potential difference $V_0$ between the free-standing region of graphene and the one under the metal:
\begin{equation}\label{eV0}
eV_{0} = \Delta E_{\rm FM} - \Delta E_{\rm FG}.
\end{equation}
This leads to a potential barrier for the electrons traveling along the $x$-direction, from one metallic contact to the other, as detailed in the next subsection. Since transmission through the potential barrier is foremost sensitive to the slope of the barrier at the charge neutrality point, we have adopted a trapezoidal barrier shape where the slope is constant on both sides of the flat  top of the barrier. This trapezoidal form is tractable using analytical calculations, and it is expected to catch the basic features of the tunneling transport problem.



\subsection{Klein tunneling: conductance and Fano factor}
\label{Klein}

The asymmetry in electrical transport across a basic, back-gated graphene device is related to many aspects of the sample, its  quality, and biasing conditions. It depends on the nature of transport, whether it is ballistic or diffusive, on the presence of interfaces between \textit{p}- and \textit{n}-doped regions, on the steepness of the slopes of the electrostatic potential barrier,  on the doping due to the contacts, and on  the coherence of the transport between the reflecting interfaces (\textit{pn}-type of interface or two unipolar regions with different doping, either \textit{p} and \textit{p'} or \textit{n} and \textit{n'}, yielding \textit{pp'}- and \textit{nn'}-interfaces). The type of junction is determined by the sign of the Fermi level shifts: if $\Delta E_{\rm FM} >0$ and $\Delta E_{\rm FG} >0$ we have a $nn'$ junction at the left contact, if $\Delta E_{\rm FM} <0$ and $\Delta E_{\rm FG} >0$ we have a $pn$ junction, if $\Delta E_{\rm FM} >0$ and $\Delta E_{\rm FG} <0$ we have a $np$ junction, and if $\Delta E_{\rm FM} <0$ and $\Delta E_{\rm FG} <0$ we have a $pp'$ junction. The same notions apply to the right contact provided that the regions are considered from right to left, instead of the left-to-right direction used for the left contact above.

Our experimental data deal with suspended graphene with a mean free path on the order of the sample length. Consequently, we will consider foremost Klein tunneling in the ballistic regime and neglect the influence of disorder \cite{Fogler2008}. Unlike the early work discussed above, we take into account the weak doping of contact regions when using Au/Cr/Au leads. We assume symmetric contacts although in reality there is always slight asymmetry. In our context, this assumption means that the values of the work function difference $\chi$ are  equal  at both contacts. The asymmetry of the contact resistance can be neglected as the contact resistance is found to be insignificant in the analysis. The use of moderate voltages and cryogenic temperatures also guarantees negligible role of acoustic phonons   \cite{DasSarma2012}.

\begin{figure}[htb]
\centering
\includegraphics[width=80mm]{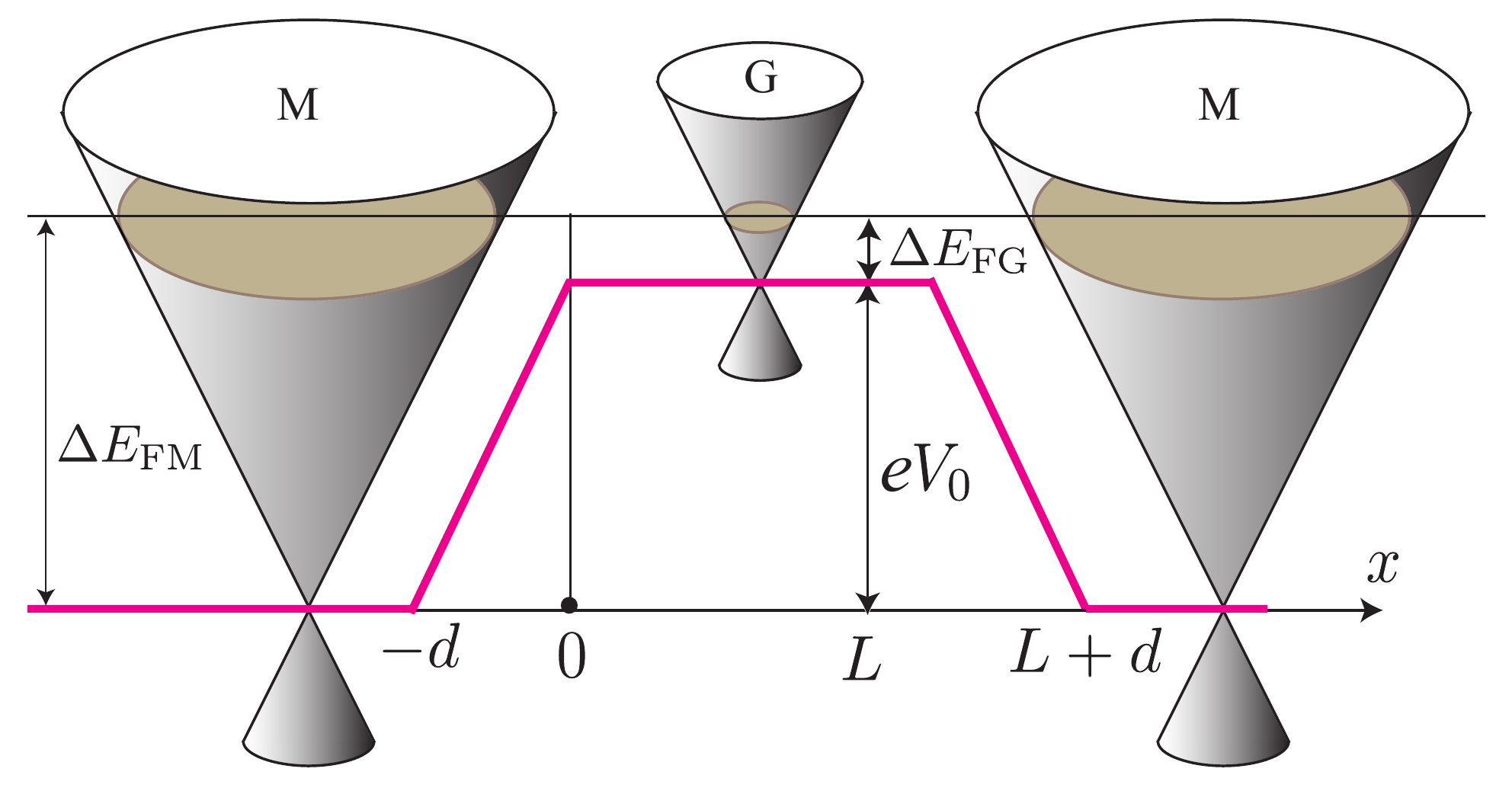}

\caption{(Color online) Schematic view of a trapezoidal \textit{nn'n}-barrier configuration across the sample. $V_0$ denotes the difference in the electrostatic potential between the contact regions and the center. The solid red line shows variation of the electrostatic potential along the graphene sheet (see also Fig.~\ref{fig1}); the kinks of the red line separate the five different spatial regions in which the Dirac equation is solved. The slope of $V(x)$ in the range $-d\leq x\leq 0$ is $a\hbar v_{\rm F}/e$.}
\label{fig2}
\end{figure}

Below, we derive the formulae for the transmission and reflection coefficients, as well as the conductance and Fano factor, for the tunneling geometry given in Fig. \ref{fig2}. The figure shows variation of the electrostatic potential (see Fig. \ref{fig1}) in the $x$-direction along the whole graphene sheet including the end contacts.  We consider only propagating modes; this approximation is justified by the criterion that propagation through the evanescent modes can be neglected if the bias voltages are clearly above $\hbar v_{F}/eL$ \cite{Sonin2008}, where $L$ is the length
of the sample. For a typical sample of length $L=1$ $\mu$m,
$\hbar v_{F}/eL=0.7$ mV, well below the bias voltages used in our  experiments for shot noise ($10 - 70$ mV).

The use of the Dirac equation with a trapezoidal barrier is the simplest approximation for transport in  a two-lead graphene sample with a pair of\textit{ pn}-interfaces due to spatially varying charge doping.
The electrostatic trapezoidal barrier (of structure \textit{nn'n}) in Fig.~\ref{fig2} is comprised of  five distinct regions: the middle part $0 \leq x \leq L$, corresponding to the region G from the previous subsection, where there is a step in electrostatic potential
$V_{0}$ (difference in the potential between the contact regions and the center), two sloped regions $-d\leq x\leq 0$ and $L\leq x \leq L+d$ in the vicinity of the graphene/metal interfaces, where the electric field
is finite, and finally the contact area $x \leq -d$ and $L+d \leq x$
 with zero (or small bias) potential, corresponding to the region M of graphene fully covered by the contact metal.
Dirac electrons propagating along the $x$-direction may thus experience Klein tunneling under this barrier.
In the left slope region $-d\leq x\leq 0$, the potential is written in the form $V(x) = V_{0} + {\rm sgn}\left[V_{0}\right]a\frac{\hbar v_{\rm F}}{e} x$. The parameter $a$ sets the absolute value of the voltage slope. In general, the barrier can have either a positive or a negative slope, depending on the sign of $V_{0}$. For clarity, we will give the explicit analytical forms of the solutions of the Dirac equation only for the case of positive slope, corresponding to the left side of the barrier in Fig.~\ref{fig2}. The solutions for negative slope can be obtained by the same procedure.

All the energies are measured from the Dirac point of the graphene under the metal, with the standard convention that $\Delta E_{\textrm{FM}}$ is positive if the Fermi level is above the Dirac point and negative otherwise, see Eqs. (\ref{fm3}) and (\ref{sol1}). A similar convention is used for $\Delta E_{\textrm{FG}}$, see Eq. (\ref{fg4}-\ref{sol2}).

The massless Dirac equation for a graphene sheet can be written as
\begin{equation} \label{D1}
\left[-i \frac{\partial}{\partial x}\sigma_{z} -i \frac{\partial}{\partial y}\sigma_{y}\right] \Psi (x,y) = {\cal K} (x) \Psi (x,y) ,
\end{equation}
where for particles near the Fermi level ${\cal K} (x) = [\Delta E_{\rm FM}- e V(x)]/\hbar v_{\rm F}$, and the upper and lower components of the spinor $\Psi (x,y)$ are denoted by $\psi_{+}(x,y)$ and $\psi_{-}(x,y)$ respectively. Note that the electrons at the Fermi level will have energy $\Delta E_{\rm FM}$ in any of the regions M, G, or in the slope region.
We can solve the Dirac equation in Eq. (\ref{D1}) by the method of separation of variables, writing $\psi_{\pm}(x,y)=\psi_{\pm}(x)\exp (ik_{y} y)$, with the current along the $x$ direction $j_{x} = ev_{\rm F}(|\psi_{+}|^2 - |\psi_{-}|^2)$ normalized to $\pm ev_{\rm F}$.

The height of the energy barrier
$eV_{0} = \Delta E_{\rm FM}- \Delta E_{\rm FG}$
defines a step $eV_{0}/\hbar v_{\rm F}$ in the momentum of the Dirac electrons as they cross the barrier. The influence of this change in momentum can be characterized compactly using
a dimensionless parameter, \textit{the impact parameter}, defined as
\begin{equation}
p_{0} = \frac{eV_{0}}{\hbar v_{\rm F}\sqrt{a}}. \label{p01}
\end{equation}
The impact parameter will play an essential role in our calculations below. Note that the earlier theoretical treatment by Sonin \cite{Sonin2009} assumed a large impact parameter $|p_{0}|\gg 1$. As this approximation is not valid for our present experiment, a finite $p_{0}$ needs to be taken into account in the theory.

 Assuming that the thickness of the sloped region $d$ is independent of the gate voltage, allows us to express the absolute value of the slope of the barrier in terms of the $V_g$-dependent quantities, $\Delta E_{\rm FM}$ and  $\Delta E_{\rm FG}$, which yields
\begin{equation}
a=\frac{\sqrt{\pi}}{\zeta_{\rm F}d}\vert V_{0}\vert.\label{aaa}
\end{equation}
Likewise, we obtain for the impact parameter
\begin{eqnarray}
p_{0}&=&{\rm sgn}\left[V_{0}\right] \left( \frac{\sqrt{\pi}d}{\zeta_{\rm F}}\right)^{1/2}\sqrt{\vert V_{0} \vert}.\label{pee}
\end{eqnarray}
Note that the impact parameter increases with the absolute value of the slope, as evident from Eqs. (\ref{p01}) and (\ref{aaa}).

\subsubsection{Top of the barrier, $x>0$ (region G)}

In this region we consider an electron moving to the right, $j_{x}=ev_{\rm F}$. The Dirac equation can be solved, with
\begin{equation}
{\cal K}(x|x>0) = \frac{1}{\hbar v_{\rm F}}\Delta E_{\rm FG},
\end{equation}
for both positive and negative $\Delta E_{\rm FG}$,
\begin{equation}
\left. \left(\begin{array}{c} \psi_{+}(x) \\ \psi_{-}(x) \end{array} \right)\right|_{x>0} = \left( \begin{array}{c} \frac{1}{2} + \frac{k_{x,{\rm FG}} \pm i k_{y}}{2k_{\rm FG}} \\ -\frac{1}{2} + \frac{k_{x, {\rm FG}} \pm i k_{y}}{2k_{\rm FG}} \end{array} \right) \sqrt{\frac{k_{\rm FG}}{k_{x, {\rm FG}}}}e^{\pm ik_{x, {\rm FG}}x},
\end{equation}
where the $\pm$ sign denotes ${\rm sgn[\Delta E_{\rm FG}]}$.
The absolute value of the wavevector is
\begin{equation}
k_{\rm FG} = \frac{\left|\Delta E_{\rm FG}\right|}{\hbar v_{\rm F}}, \label{kFG}
\end{equation}
 and $k_{x, {\rm FG}} = \sqrt{k_{\rm FG}^2-k_{y}^2}$. Note that $k_{y}$ remains unchanged over all region crossings.

\subsubsection{Slope of the barrier, $-d<x<0$}

In this region we have
\begin{equation}
{\cal K}(x|-d<x<0) = -{\rm sgn}\left[V_{0}\right] a(x-x_{0}),
\end{equation}
depending on weather the slope is positive or negative. Here $x
_{0}$ is the crossing point, defined as the position where the kernel is nullified, and it
is given by $x_{0}= {\rm sgn}\left[V_{0}\right]\Delta E_{\rm FG}/a\hbar v_{\rm F}$.
The solutions have been found by Sauter (see Ref. \onlinecite{Sonin2009}); for positive slope, with the substitution $\xi (x) = \sqrt{a} (x-x_{0})$
we have
\begin{eqnarray}
\left.\psi_{+}(x)\right|_{-d<x<0}
&=&C_{1} F(\xi (x), k_{y}/\sqrt{a}) +  C_{2} G^{*}(\xi (x), k_{y}/\sqrt{a}) \nonumber\\
\left.\psi_{-}(x)\right|_{-d<x<0} &=&C_1 G(\xi (x) , k_{y}/\sqrt{a}) + C_{2} F^{*}(\xi (x), k_{y}/\sqrt{a}). \nonumber
 \end{eqnarray}
 Note also that the dimension of $a$ is [length]$^{-2}$, while that of ${\cal K}(x)$ and $k_{y}$ is [length]$^{-1}$; hence  the quantity $k_{y}/\sqrt{a}$, that enters the hypergeometric functions, is adimensional.

The functions $F$ and $G$ are defined through the Kummer confluent hypergeometric function $M\equiv~_{1}F_{1}$,
\begin{equation}
F(\xi, \kappa) = e^{-i\xi^2 /2}M\left(-\frac{i\kappa^2}{4}, \frac{1}{2}, i\xi^2\right),
\end{equation}
and
\begin{equation}
G(\xi, \kappa) = - \kappa \xi e^{-i\xi^2/2}M\left(1- \frac{i\kappa^2}{4}, \frac{3}{2}, i\xi^2 \right).
\end{equation}


From the boundary conditions at $x=0$ we obtain
\begin{widetext}
\begin{eqnarray}
C_{1} &=& \frac{k_{\rm FG} + k_{x, {\rm FG}} + i {\rm sgn}[\Delta E_{\rm FG}] k_{y}
}{2\sqrt{k_{\rm FG}k_{x, {\rm FG}}}}F^{*}(\xi (0), k_{y}/\sqrt{a})+
\frac{k_{\rm FG} - k_{x, {\rm FG}} - i {\rm sgn}[\Delta E_{\rm FG}] k_{y}
}{2\sqrt{k_{\rm FG}k_{x, {\rm FG}}}}G^{*}(\xi (0), k_{y}/\sqrt{a}) , \\
C_{2} &=& \frac{-k_{\rm FG} + k_{x, {\rm FG}} + i {\rm sgn}[\Delta E_{\rm FG}] k_{y}
}{2\sqrt{k_{\rm FG}k_{x, {\rm FG}}}}F(\xi (0), k_{y}/\sqrt{a})-
\frac{k_{\rm FG} + k_{x, {\rm FG}} + i {\rm sgn}[\Delta E_{\rm FG}] k_{y}
}{2\sqrt{k_{\rm FG}k_{x, {\rm FG}}}}G(\xi (0), k_{y}/\sqrt{a}) .
\end{eqnarray}
\end{widetext}
Importantly, a consequence of the normalization $|\psi_{+}|^2 - |\psi_{-}|^2=1$ is the fact that $|C_{1}|^2
-|C_{2}|^2 = 1$ and $|F(\xi, \kappa)|^2 - |G(\xi, \kappa)|^2 =1$, which can be verified explicitly using the explicit expressions above.

\subsubsection{Graphene under the metal $x<-d$ (region M)}

In the calculation of Sonin \cite{Sonin2009}, this region with $x < -d$ was disregarded on the basis of the assumption $p_0^2 \gg 1$). It turns out that this condition is not valid in our experiment (cf. Fig. \ref{fig:QV}), and the behavior in the region $x < -d$ has to be taken into account.
In this region,
\begin{equation}
{\cal K}(x|-d<x<0)  = \frac{1}{\hbar v_{\rm F}}\Delta E_{\rm FM}.
\end{equation}
The absolute value of the total momentum at the Fermi level is then
 \begin{equation}
  k_{\rm FM} = \frac{1}{\hbar v_{\rm F}}\vert \Delta E_{\rm FM}\vert .\label{kFM}
 \end{equation}
 The corresponding momentum in the x-direction equals to
 $k_{x, {\rm FM}} =\sqrt{k_{\rm FM}^2-k_{y}^2}$,
and the wave is a superposition of a reflected and a transmitted component,
\begin{widetext}
\begin{equation}
\left. \left(\begin{array}{c} \psi_{+}(x) \\ \psi_{-}(x) \end{array} \right)\right|_{x<-d} = \frac{1}{t} \left(\begin{array}{c} \frac{1}{2} + \frac{k_{x,\rm FM} \pm i k_{y}}{2k_{\rm FM}} \\ -\frac{1}{2} +
\frac{k_{x,\rm FM} \pm
i k_{y}}{2k_{\rm FM}} \end{array}\right) \sqrt{\frac{k_{\rm FM}}{k_{x,\rm FM}}}e^{\pm i k_{x,\rm FM}x} +
\frac{r}{t}\left(\begin{array}{c} \frac{1}{2} + \frac{-k_{x,\rm FM} \pm i k_{y}}{2 k_{\rm FM}} \\ -\frac{1}{2} + \frac{-k_{x,\rm FM} \pm ik_{y}}{2 k_{\rm FM}}\end{array}\right) \sqrt{\frac{k_{\rm FM}
}{k_{x,\rm FM}}}e^{\mp i k_{x,\rm FM}x}. \label{admixt}
\end{equation}
\end{widetext}
Here $r$ and $t$ are the reflection and transmission amplitudes, $\pm$ is the sign of ${\rm sgn}[\Delta E_{\rm FM}]$, and $\mp$ in the last exponent is $ - {\rm
sgn}[\Delta E_{\rm FM}]$.

Note also that in Eq. (\ref{admixt}) the reflected component is normalized to the current in the x-direction, equal to $-ev_{\rm F}$, while the transmitted component is normalized to $+ev_{\rm F}$. However, one can explicitly check that the overall normalization of $\Psi (x,y)$ is to $+ev_{\rm F}$, provided that $|r|^2 + |t|^2=1$. This ensures that the normalization
is the same for all the three regions. To understand intuitively how this is realized, note that the admixture of the reflection component in $\Psi (x,y)$ of Eq. (\ref{admixt}) is compensated by an increase in the component propagating to the right, since $|t|$ becomes subunitary.

By imposing the condition of continuity of the wave function at $x=-d$, after some algebra we obtain the complex transmission amplitude

\begin{widetext}
\begin{eqnarray}
\frac{1}{t}e^{-i k_{x, {\rm FM}}d} &=&
-\frac{-k_{\rm FM} - k_{x, {\rm FM}} + i {\rm sgn}[\Delta E_{\rm FM}] k_{y}
}{2\sqrt{k_{\rm FM}k_{x, {\rm FM}}}}[C_{1} F(\xi (-d), k_{y}/\sqrt{a}) + C_{2} G^{*}(\xi (-d), k_{y}/\sqrt{a})] \nonumber \\ &  & +
\frac{k_{\rm FM} - k_{x, {\rm FM}} + i {\rm sgn}[\Delta E_{\rm FM}] k_{y}
}{2\sqrt{k_{\rm FM}k_{x, {\rm FM}}}}[C_{1} G(\xi (-d), k_{y}/\sqrt{a}) + C_{2} F^{*}(\xi (-d), k_{y}/\sqrt{a})]. \label{transmission}\end{eqnarray}
\end{widetext}


\subsection{Conductance and Fano factor for the whole barrier} \label{incoh_arg}

To calculate the total transmission through the barrier, we employ incoherent addition of the transmission coefficients.
Usually phase coherence is more sensitive to disorder than reflection and transmission, and we address the case when the former is destroyed but the latter is not affected by disorder (ballistic regime).
This assumption is supported by the experimental fact that the \FP resonances are found to be weak.  Furthermore, when destroying the \FP resonances fully by an applied bias, the overall conductance does not change much. Hence, we consider the incoherent treatment of  transmission probabilities well justified in our analysis.

In general, for the case of incoherent tunneling through a symmetric barrier with equal transmission $T$ for the left and right slopes, the total probability of transmission through the barrier is (see Appendix \ref{incoherent})
\begin{equation}
T_{\rm tot}= \frac{1}{2 T^{-1}-1}. \label{totaltransmission}
\end{equation}
By using the Landauer-B\"uttiker formalism, we obtain the conductance and the Fano factor as sums over the transmission coefficients and their quadratic values.
Each quantized value of $k_{y}$ corresponds to a conduction channel, over which the summation of transmission coefficients has to be performed in order to obtain the total conductance (shot noise) from the conductance per channel (shot noise per channel).  Thus, for a sample with a given level of contact doping $\Delta E_{{\rm FM}}$, one can calculate the conductance $\sigma$ and the Fano factor $F$ as a function of gate voltage. In the limit of large number of channels, $\sigma$ and $F$ can be written as
\begin{eqnarray}
\sigma  &=& \frac{4e^{2}W}{\pi h}\int_{0}^{{\rm min}\left\{k_{\rm FM}, k_{\rm FG}\right\}}
d k_{y} T_{\rm tot}, \label{gg}\\
F &=& 1- \frac{1}{\sigma}\int_{0}^{{\rm min}\left\{k_{\rm FM}, k_{\rm FG}\right\}}d k_{y} T_{\rm tot}^{2}. \label{FF}
\end{eqnarray}
The conductance and the Fano factor will depend on $V_{g}$ through the dependence of the quantities $k_{\rm FG}$, $k_{\rm FM}$, $\Delta E_{\rm FG}$, $\Delta E_{\rm FM}$ obtained previously.
If the contact resistance between the graphene and the metal is neglected, then the input to Eq. (\ref{totaltransmission}) is given by $T= |t|^2$, where $|t|^2=|t(k_{y})|^2$ is the transmission coefficient calculated using Eq. (\ref{transmission}).

If a finite contact resistance exists, then a finite transmission probability
$0 \leq T_{c} \leq 1$ should be included in the value of $T$ in Eq. (\ref{totaltransmission}). By applying again the result of Appendix \ref{incoherent}, we may write
\begin{equation}
T = \frac{T_{c}|t|^2}{T_{c}+|t|^2-T_{c}|t|^2}. \label{contactresistance}
\end{equation}
The inclusion of contact resistance has a strong influence on the shot noise, and  the calculated Fano factor becomes quickly larger than the measured value when the contact transmission is lowered from one.

The limits of integration in Eqs. (\ref{gg}) and (\ref{FF}) are set by the condition that the wave vector $k_{x}$ is a real number, in other words the electron is not in a bound state but propagates to infinity. The condition
$k_{y} <k_{\rm FG}$ comes from the top of the barrier region, while the condition $k_{y} <k_{\rm FM}$ comes from the region $x<-d$.

Interestingly, even though our potential profile with five distinct regions as shown in Fig.~\ref{fig2}, the entire model has only two essential fitting parameters, namely $p_0=eV_{0}/\hbar v_{\rm F}\sqrt{a}= {\rm sgn}\left[V_{0}\right]\sqrt{a}d$ and the barrier slope $a$. The impact  parameter contains the information on the doping via Eqs. (\ref{fm3}), (\ref{sol2}), and (\ref{eV0})  and it enters in the upper limit of the integrals in Eqs. (\ref{gg}-\ref{FF}), since $k_{\rm FM} = |\sqrt{a}p_{0} + \Delta E_{\rm FG}/\hbar v_{\rm F}|$.
In fact, the Fano factor in our calculation is fully determined by the value of $p_0$, while the conductance integrals need also the value of the barrier slope for their evaluation.

\section{Sample and shot noise methods} \label{samplenoise}

\begin{figure}
\centering

\includegraphics[width=70mm]{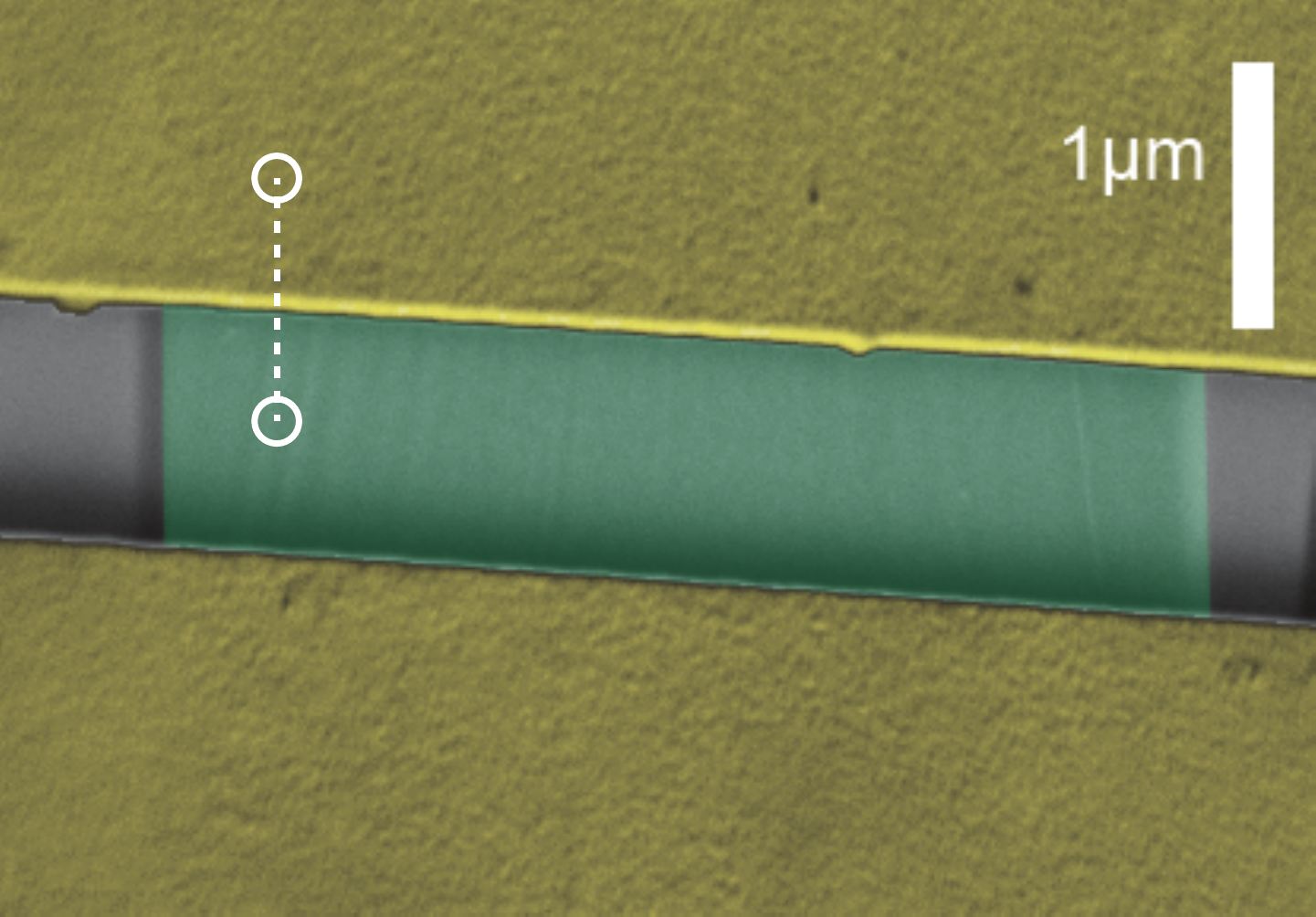}
\caption{(Color online)  False color scanning electron micrograph  taken of a suspended sample similar to the measured one (topview on y-x plane; x-axis is oriented vertically): yellowish areas denote Au/Cr/Au contacts while the greenish area in between depicts the suspended part of the graphene sheet. Wrinkles due to strain are visible in the sheet at room temperature. Charge is induced into the central graphene region by the backgate (formed by the doped silicon substrate, not visible) and, consequently, a pair  of junctions (abrupt changes in the charge density) are formed very close to the  contacts. The upper white circle denotes a position where the schematic profiles of Fig. 1b apply, while the profile at the lower circle is governed by Fig. 1c.  The scale is given by the white bar on the right.}
\label{schem}
\end{figure}

The measured suspended sample was manufactured using standard PMMA-based e-beam lithographic techniques on a graphene piece exfoliated onto Si/SiO$_2$ chip (see Fig.~\ref{schem}). The dashed white line with two circles in the figure indicates the positions at which one obtains the cross sectional view depicted in Fig.  \ref{fig1}a; the circles correspond to the spots at which the electrochemical potential profiles (along vertical direction) are drawn in Figs. \ref{fig1}b and 1c.   The graphene sample was contacted with Au/Cr/Au leads of thickness 5/7/50 nm, after which roughly 135 nm of the sacrificial silicon dioxide was etched away using hydrofluoric acid (HF) following the methods discussed in Refs. \onlinecite{Khodkov2012} and \onlinecite{Khodkov2015}. Raman spectroscopy was employed to verify the single layer structure of the graphene sample. Current annealing at $V_g = 0$ was used to enhance the mobility of the sample. We employed  voltage bias around 1 V and a current of $0.3-0.7$ $\mu$A/$\mu$m in our cleaning process. The aspect ratio of the sample as determined before the experiments  was $W/L = 4.5$ $\mu {\rm m} /1.1$ $ \mu {\rm m} \approx 4.1$. The capacitance $C_{\rm gG} = 47$ aF/$\mu$m$^2$ was determined  using \FP interference fringe measurements \cite{Oksanen2014}. The mobility of the sample $\mu_{\rm G}$ was calculated using the charge carrier density  $|n_{\rm G}|$ = $|C_{g{\rm G}}\delta V_g|/e$ (obtained from Eq. (\ref{sol2}) in the limit of large $\delta V_g$) and $\sigma(\delta V_g) = R^{-1}(\delta V_g)L/W$ in the formula  $\mu_{\rm G} = (\sigma-\sigma_{\rm min}^{\rm meas})/n_{\rm G}e$, where $\sigma_{\rm min}^{\rm meas}$ is the measured minimum conductivity. We find $\mu_{\rm G} >10^5$ cm$^2$/Vs near the Dirac point at $n < 2.5 \cdot 10^{10}$ cm$^{-2}$. For the Fermi velocity we used the value $v_F = 1.1 \cdot 10^6$ m/s; note that owing to interaction effects at small charge density, the Fermi velocity can grow up to $v_F \simeq 3 \cdot 10^6$ m/s in our sample near the Dirac point \cite{Oksanen2014}. The assumption of symmetric contact capacitances (within $ \pm15$ \%) was verified from the inclination of the \FP  pattern (see Ref. \onlinecite{Oksanen2014}).

Differential conductance of the sample was measured using standard low-frequency lock-in techniques (around 35 Hz); the same excitation
was also employed in our differential shot noise measurements.
The noise-signal from the sample was led via a circulator to a cryogenic low noise pre-amplifier having a bandwidth of  600-900 MHz \cite{Roschier2004}. The amplifier provided a gain of 15 dB and the signal was further amplified at room temperature by 80 dB and band pass filtered before the Schottky diode detection. Small bonding pads of size 90 $\mu$m $\times$ 90 $\mu$m were employed in order to keep the shunting capacitance $\sim 0.1$ pF negligible in our microwave noise measurements \cite{Danneau2008b}. For calibration purposes, a microwave switch was used to select a tunnel junction as the noise source instead of the sample. For details of the calibration procedure, see Ref. \onlinecite{Danneau2008b}. In our experiments, we employ the excess Fano factor obtained from the noise: $F=(S_I(V_b)-S_I(0))/2eI$, where we use the difference of the current noise spectrum $S_I$ between bias voltage $V_b=22$ mV and $V_b=0$.
The experiments were performed around 0.5 K in a Bluefors BF-LD250 dilution refrigerator.

\section{Measurements and Results} \label{experiment}

Before presenting our data and their analysis, let us first summarize  the free parameters that enter our theoretical model. The capacitances are fixed by the geometry and the charge dipole layer thickness  $z_d$, which equals to $z_d=0.9$ {\AA}   according to the calculation of Ref. \onlinecite{giovannetti2008} for graphene/gold interface; assuming $z_d$ corresponds to $d_c$ in Fig. 1, this yields $C_c = 9.8 \times 10^{-6}$ F/cm$^{2}$ using the vacuum permittivity. There are three parameters that are fitted to the data: $\chi = W_{\rm G}+\Delta_{c} - W_{\rm M}$, $ \Delta E_{\rm FM}$, and the thickness $d$ of the charge gradient interface. The overall magnitude of the measured conductance sets $\chi=- 18$ meV in our analysis. Furthermore, the noise analysis indicates that $d \sim 20$ nm (see below). Hence, we are left with one $\delta V_g$-dependent fit parameter, $\Delta E_{\rm FM}$, which describes the electrostatic behavior of the contact regime with the gate voltage.

\begin{figure}[H]
\centering
\includegraphics[width=80mm]{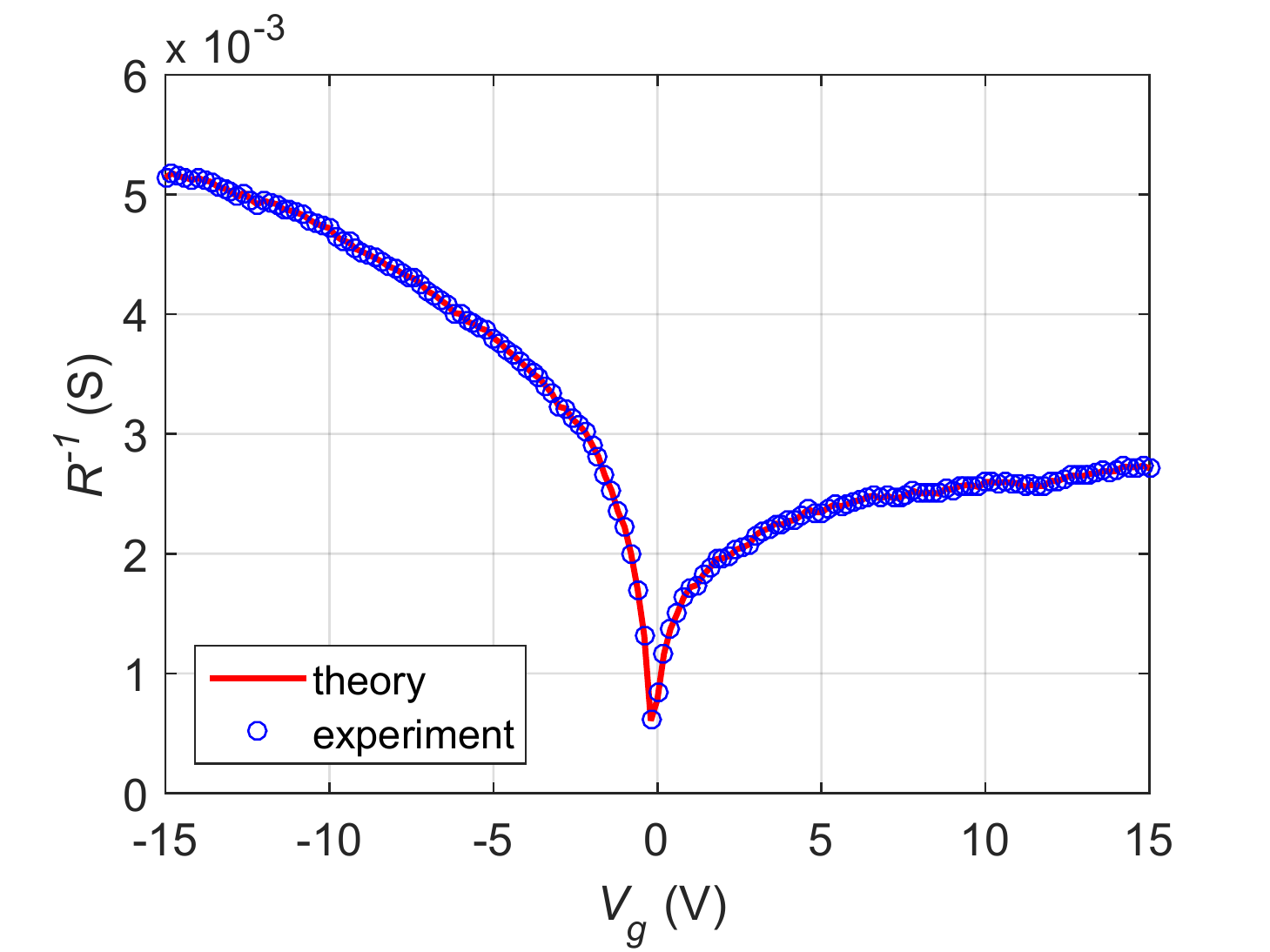}
\caption{(Color online) Zero-bias conductance $R^{-1}$ as a function of the gate voltage $V_g$; $V_g > V_{\rm Dirac}$ corresponds to a $pnp$ structure, while $V_g > V_{\rm Dirac}$ represents data in a $pp'p$ configuration. The overlaid trace depicts the curve which was employed to determine the Fermi level shift $\Delta E_{\textrm{FM}}(V_g)$; this  Fermi level shift is displayed \emph{vs. }$\delta V_g= V_g-V_{\textrm{Dirac}}=V_g+0.2$ V in Fig. \ref{fig:QV}. }
\label{fig:R_pn}
\end{figure}

The zero-bias conductance $\sigma$ as a function of  $V_g$ is displayed in Fig. \ref{fig:R_pn}. The Dirac point resides at $V_{\rm Dirac}$ = $-0.2$ V indicating negative dopants on the sample. However, the asymmetry of the conductance and the shot noise suggest positive doping, which would correspond to $V_{\rm Dirac} > 0$ V, in contrast to the above. In order to account for the "wrong" sign of the Dirac point location, we argue that the work function of the back gate has to play a role. The work function of doped silicon depends on the sign and amount of the dopants. At large \textit{negative} doping, $W_g  \simeq 4.4$ eV for our background material with a negative dopant concentration of $\sim 10^{16}$ cm$^{-2}$ \cite{novikov2010}.  Since $W_G = 4.6$ eV according to Ref. \onlinecite{nagashio1997}, we obtain $V_{\textrm{Dirac}}=W_g - W_G =-0.2$ V, which leads to $\delta V_g= 0$ at $V_g= -0.2$ V. Hence our data are consistent with doping induced by the work function of the backgate material.

 As in earlier works, we characterize our results in terms of the odd part of the resistance across the Dirac point, $R_{\textrm{odd}}=(R(+\delta V_g)-R(-\delta V_g))/2$ with $\delta V_g$ counted from the Dirac point, we find $R_{\textrm{odd}}=100-70$ $\Omega$ at carrier density $|n_{\rm G}|=0.3-4 \cdot 10^{11}$ cm$^{-2}$, which is consistent with other experiments \cite{Huard2007, Huard2008, Stander2009}. The conductance at large positive $\delta V_g$ is substantially smaller than at equal carrier density at $\delta V_g <0$ as expected in the presence of $pn$-interfaces.
The overlaid curves in Fig. \ref{fig:R_pn} represent the fits of the model for each gate voltage value using the Fermi level position $ \Delta E_{\textrm{FM}}(V_g)$ as the fitting parameter, with set values for the thickness $d=20$ nm and for the work function difference $\chi  = -18$ meV; further input parameters of the model are specified in Sect. \ref{parameters}. Since the fitting is done separately at each point, there is no difference between the overlaid curve and the data.

\begin{figure}[H]
\centering
\includegraphics[width=80mm]{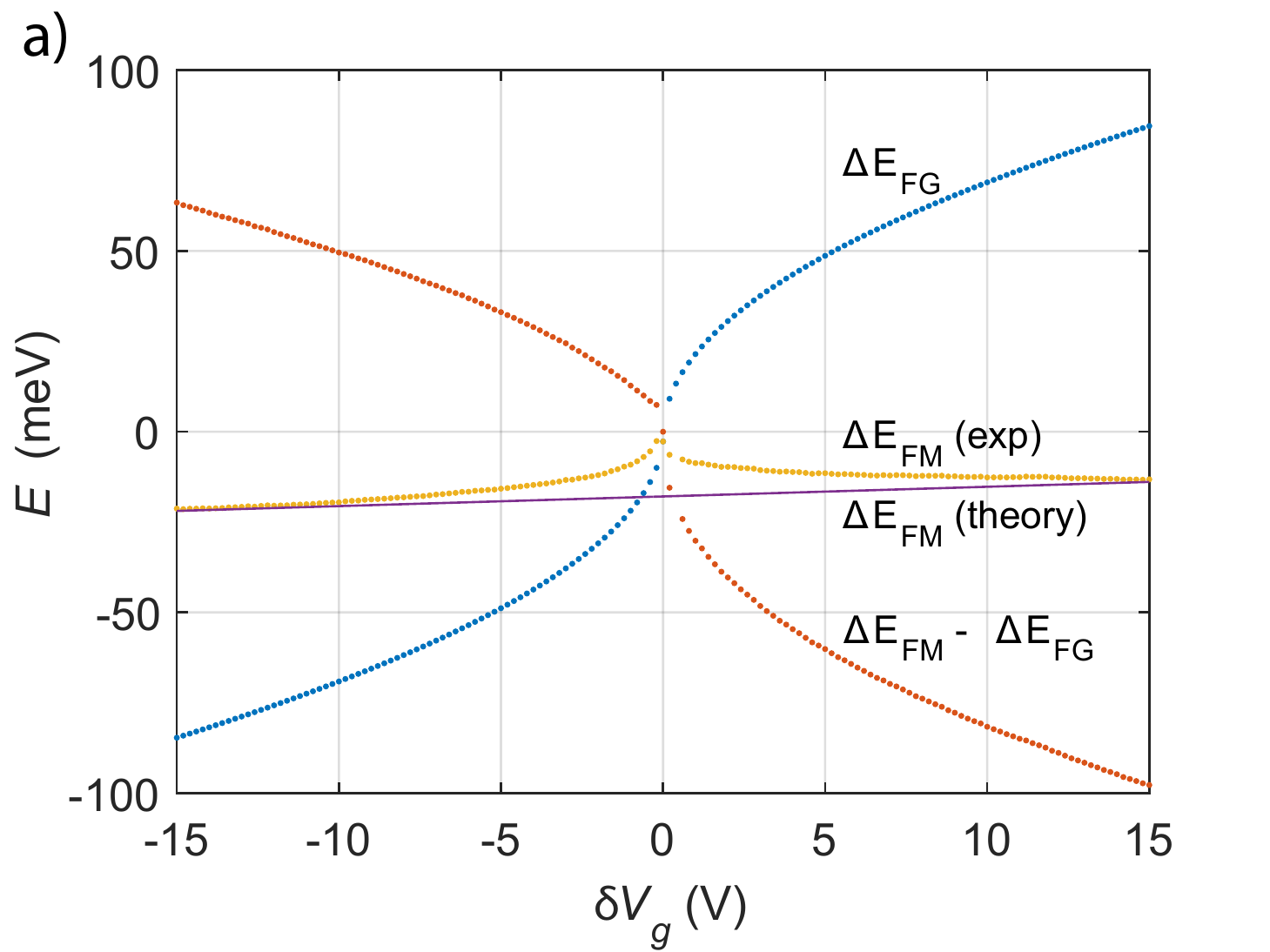}
\includegraphics[width=80mm]{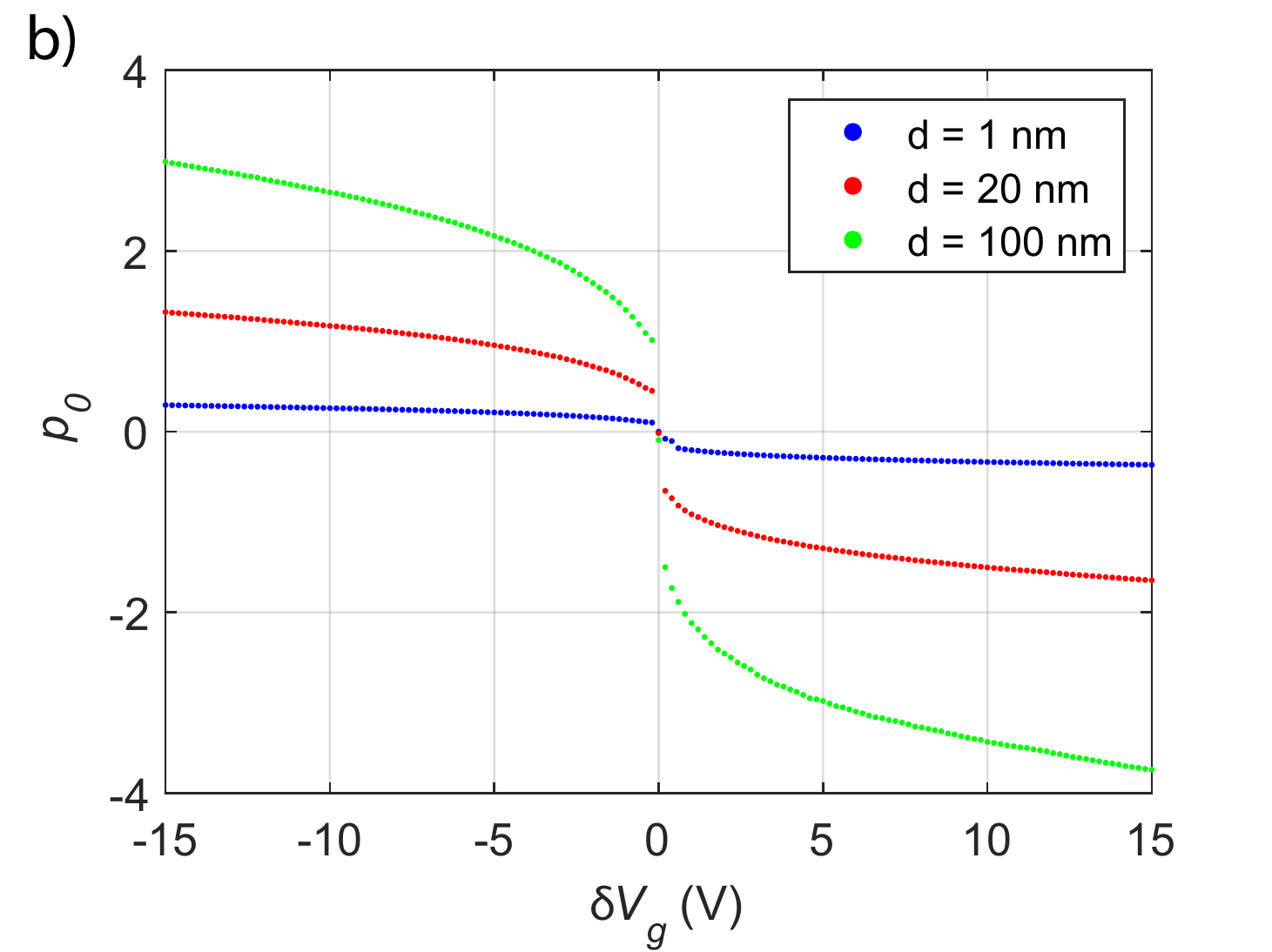}
\caption{(Color online) (Upper frame) Position of the Fermi level $ \Delta E_{\textrm{FM}}$ in the contact region obtained from the conductance analysis of Fig. \ref{fig:R_pn} (orange symbols), the Fermi level  position in the suspended part $\Delta E_{\textrm{FG}}$ (blue symbols), and the difference $ \Delta E_{\textrm{FM}} -  \Delta E_{\textrm{FG}}$ (red symbols). The solid line is the dependence obtained from Eq. (\ref{sol1}) using $\chi = -18$ meV; the values of the contact capacitances are discussed in the text. (Lower frame) Variation of the impact parameter $p_0$  with gate voltage  obtained from the  $ \Delta E_{\textrm{FM}} -  \Delta E_{\textrm{FG}}$ data presented above; $p_0 < 0$ corresponds to the $ pnp$ configuration. The behavior depends on the thickness $d$, the values of which are indicated in the figure.
 }
\label{fig:QV}
\end{figure}

Fig. \ref{fig:QV}a depicts the Fermi level shift $\Delta E_{\textrm{FM}}$ in the contact regions, obtained from the point-wise fitting of our model to the conductance data in Fig. \ref{fig:R_pn}. The Fermi level does not move much with $\delta V_g$, which is due to a large contact capacitance $C_c$ when compared with the other capacitances  in the system (see Eq. (\ref{sol1})).  Neglecting the region around the Dirac point, $\Delta E_{\textrm{FM}}$  changes almost linearly with gate voltage due to the additional charge induced by $V_g$ to the contact region. This is expected from our model, and by fitting Eq. (\ref{sol1}) to our data in Fig. \ref{fig:QV}a we obtain $\chi = -18$ meV, $C_c=1.2 \times 10^{-5}$ F/cm$^2$, and $C_{g{\rm M}} = 6 \times 10^{-9}$ F/cm$^2$, close to the geometrical estimates discussed earlier. Taking into account the uncertainties in the involved capacitances, the asymptotic agreement between the experiment and the theoretical model can be considered as good.

Additional effects are observed near the Dirac point $\delta V_g = 0$.
The doping $\Delta E_{\rm FM}$ becomes close to zero, and clearly our electrostatic model does not capture  the behavior completely.  One shortcoming, for example, is that we do not include  the renormalization of Fermi velocity by interactions near the Dirac point,  which would partly  improve the agreement between Eq. (\ref{sol1}) and the measured data. Also near the Dirac point evanescent waves become important, but they can be neglected when working at finite $V_{b} \gg 2 \hbar v_F/eL$.  Finally, another explanation could be the formation of charge puddles, which exist in suspended graphene even though their strength is suppressed compared with non-suspended samples \cite{Du2008}. These phenomena are beyond our present analysis.

Fig. \ref{fig:QV}a displays our experimental results for $\Delta E_{\textrm{FG}}$ and the difference $\Delta E_{\textrm{FM}} -  \Delta E_{\textrm{FG}}$. Once $ \Delta E_{\textrm{FM}} -  \Delta E_{\textrm{FG}}$ is known, we may evaluate the impact parameter $p_{0}=eV_{0}/\hbar v_{\rm F}\sqrt{a}$, which is  depicted in Fig. \ref{fig:QV}b. We note that the value of the impact parameter is proportional to $\sqrt{d}$ (when $V_0$ is fixed, $a \propto 1/d$), apart from small corrections.
At large interface thickness, our data approaches the regime of validity of the analysis by Sonin \cite{Sonin2009}.
According to Eq. (\ref{aaa}), the $\delta V_g$ dependence of the slope of $a$ in our barrier is given by $\Delta E_{\textrm{FM}} -  \Delta E_{\textrm{FG}}$ when the interface thickness $d$ is fixed.

\begin{figure}[H]
\centering
\includegraphics[width=80mm]{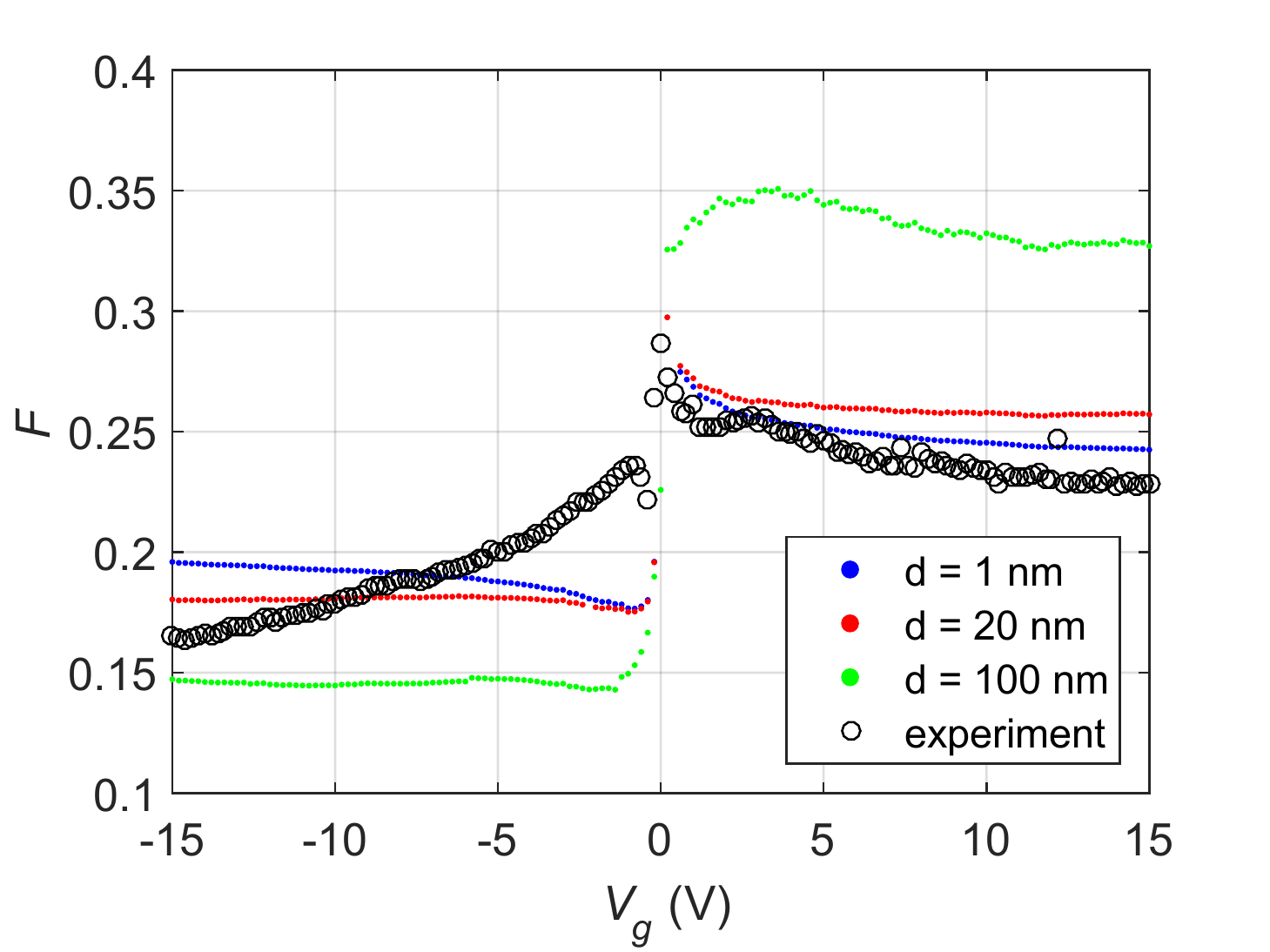}
\caption{(Color online) Measured Fano factor (circles) vs gate voltage$V_g$, and  results from the contact model using parameters obtained from the conductance; the curves are calculated for thickness $d=1$, 20, and 100 nm (blue, red, green, respectively).}
\label{fig:mult}
\end{figure}

Last, we present our data on the Fano factor.  We employed a bias voltage $V_b > 20$ mV in our measurements in order to be well above the cross-over voltage between thermal and shot noise, as well as to reach clearly the regime where the incoherent summation of barrier transmissions becomes valid (see Subsection \ref{incoh_arg}). The experimental data measured at $V_{b}=22$ mV are presented in Fig. \ref{fig:mult} together with the curves given by our theoretical formula in  Eq. (\ref{FF}) for interface thicknesses $d=1$, 20, and 100 nm between the doping sections dominated by the metal and the backgate. The dependence of the calculated Fano factor with the interface thickness is found to be rather weak below $d=20$ nm.

As with resistance, we have analyzed the odd part of the Fano factor, $F_{\textrm{odd}}=F(+\delta V_g)-F(-\delta V_g)$ , which would be independent of spurious, common contributions due to scattering on both sides of the Dirac point.  For example at $\delta V_{g}=72$ mV, we find an asymmetric maximum at $|k| \simeq 3 \cdot 10^5$ cm$^{-1}$ above which $F_{\textrm{odd}}$ decreases smoothly from  $\simeq 0.1$ down to 0.03 at charge density of $|n_{\rm G}|= 4 \cdot  10^{11}$ cm$^{-2}$. The experimentally obtained value $F_{\textrm{odd}}\simeq 0.1$ and the slow decrease with $|k|$ correspond well to the theoretical behavior.

Our noise data clearly display asymmetry that is expected for Klein tunneling. Our model yields Fano factors that are in good agreement with the measured results at $\delta V_g > 0$ (the case with $pn$-interfaces) when the interface thickness is set to 1 nm. Thick interfaces on the order of 100 nm do not fit the results whereas intermediate thicknesses $d \sim 20$ nm and below fit reasonably well. Consequently, we conclude that $d < 20$ nm on the basis of our shot noise results.

At $\delta V_g < 0$, our model predicts a clear drop in the Fano factor with decreasing $\delta V_g$ and then a subsequent gradual recovery with lowering electrochemical potential. The data displays similar features but the $\delta V_g$ scale appears much smaller.  The calculated Fano factor at $\delta V_g < 0$ appears to be  in agreement with the data on average . The data display a clear dip on the negative gate voltage side. Asymptotically, there is a clear difference in the calculated Fano factors, as in the experimental results. The agreement could be improved by adding a fitting parameter to account for a small reduction of contact transparency, but here we have set the contact transparency $T=1$ in order to avoid an additional fit parameter. We believe that these differences are mostly caused by non linear screening effects which are known to be different at the Dirac point and away from it \cite{Zhang2008,Khomyakov2010}. Such charge density dependent screening is not included in our model.

\section{Discussion} \label{disc}

In the experiments on graphene, it is common that the \FP type interferences remain weak  even though the sample is more or less ballistic \cite{Oksanen2014}. This is accordance with the starting point of our analysis in which the interference effects are neglected. Indeed, phase coherence is satisfied over small distances on the order of 20 nm (i.e. over the thickness of the barrier) as assumed in our calculations. However, the coherence is almost lost over the  distance  between the barriers and interferences become negligible.
Besides, our noise experiments were performed at finite bias, above the regime where \FP oscillations were observed, which may strengthen the tendency towards incoherent behavior; in fact,  electron-electron interaction effects in graphene have been found to be 100 larger in graphene than in regular metallic systems \cite{Voutilainen2011}.

The work function difference including the chemical interaction $\Delta_c$ was found to be related to the overall magnitude of the sample conductance, this did set $\chi=W_{\rm G}-W_{\rm M}+\Delta_{c} = -18$ meV. In our notation, $\chi < 0$ corresponds to positive doping if $\Delta E_{\rm FM}$ is solely governed by $\chi$.
For gold contacts ($W_{\rm M}=4.7$ eV), DFT calculations predict $\chi \simeq -0.1...-0.2$ eV \cite{giovannetti2008,Khomyakov2009}, but one has to keep in mind that these calculations yield the work function of pure gold with an error on the order of 0.2 eV. Experiments suggest that $\chi=-0.35$ eV for pure gold \cite{Malec2011}. Furthermore, Cr has a work function of $W_{\rm M}=4.5$ eV and Cr/Au contacts have been shown to yield a work function of 4.3 eV for graphene under the contact \cite{Song2012c}, which leads to an estimate $\chi \simeq -0.2$ eV for Cr/Au contacts.
The results of  Ref. \onlinecite{Nagashio2009} suggest $\chi = -0.1$ eV for Au/Cr contact, although  the authors discuss that the contact doping might be close to zero, which would be in agreement with our results. However, the exact comparison with  Ref. \onlinecite{Nagashio2009} is  problematic because of the differences in the contact structure: we evaporate first 5 nm of Au before laying down 7 nm of Cr. As discussed below, the vacancy creation by the first evaporated metal layer is of major concern.

The actual electrical contacting is further complicated by the reactivity of metal atoms on top of graphene \cite{zan2011,ramasse2012}. Using in situ X-ray photoelectron spectroscopy it has been demonstrated that sometimes the side-contacting picture may be misleading with real contacts \cite{gong2014}. In our analysis, for simplicity, we need to assume a uniform graphene layer under the metallic contact, although it is possible that metals like Cr promote vacancy formation and lead to creation of defects under the evaporated metal; similar defect formation has been reported due to deposition of gold atoms \cite{shen2013}, as well as in annealing studies of metallic contacts \cite{leong2014}.  Furthermore,
charge transfer  at the interface depends  on the amount of oxygen and nitrogen on top of graphene as has been found in functionalization studies by Foley \textit{et al }\cite{foley2015}.

Despite of possible defect formation, gold contacts are known to preserve the graphene cone structure under the contact \cite{Sundaram2011,Ifuku2013}. Quantum capacitance measurements of graphene under the contact were performed in Ref. \onlinecite{Ifuku2013} to characterize the cone structures. The accuracy of these quantum capacitance determinations compared with theory is within a factor of two. These findings suggest a modification of the Fermi velocity under the contacts. An inclusion of such a modification would form a promising extension of our theory, but this was left for future.

The parameter $\Delta E_{\textrm{FM}}$ together with gate capacitance (yielding $ \Delta E_{\textrm{FG}}$) determines fully the product $da$ which varies strongly with the gate voltage.  Hence, our measurement imposes a constraint on the product   $da$ as a function of $\delta V_g$. When the thickness of the interface is fixed, the slope of the interface approaches quickly zero as $\Delta E_{\textrm{FM}} - \Delta E_{\textrm{FG}} \rightarrow 0$. Altogether, the range of variation of $ \Delta E_{\textrm{FM}} - \Delta E_{\textrm{FG}}$ is rather limited, which indicates weak doping by contacting metal as well as by the gate ($C_{g{\rm G}}$ is small for suspended devices).

The same analysis as presented here can be performed using a constant slope $a$ rather than a fixed $d$. We did such an analysis and, somewhat surprisingly, the numerical simulations yielded similar predictions for the conductivity and Fano factor, in particular for $R_{\rm odd}$ and $F_{\rm odd}$. This increases the confidence in the analysis of our results, demonstrating that the extraction of relevant parameters does not depend strongly on model-specific assumptions about the trapezoidal barrier in the regime of our data (\textit{i.e}. at small interface thickness $d$).

 Screening influences the speed at which charge density varies across borders between differently doped regions. Due to its peculiar inherent properties, screening in graphene may be strongly non-linear.  According to theory \cite{Khomyakov2010}, the screening influences mostly the asymptotic behavior of the barrier, whether it is $x^{-1/2}$ or $x^{-1}$, but not much the length scale of the rapid initial relaxation. Since we use trapezoidal shape and neglect the relaxation behavior in the barrier altogether, we have chosen to work with fixed thickness, which is proportional to the average inverse slope of the charge relaxation at the $pn$-interface. One should keep in mind that our analysis is not reliable close to the Dirac point since there the slope itself weakly affects the  conductance. Therefore, our fitting is mostly sensitive to the interfacial thickness far away from the Dirac point where the screening length becomes short.

 According to Ref. \onlinecite{Khomyakov2010}, the approximate thickness of the $pn$-interface could be in the range of 5 nm (see also Ref. \onlinecite{Barraza2010}), which is consistent with our results $d< 20$ nm.  Full numerical simulations on the $pn$-interface structure in a double-gated graphene structure have been performed in Ref. \onlinecite{liu2013}. Within the trapezoidal approximation, the calculated slope of the potential profile in Ref. \onlinecite{liu2013} yields an interface thickness of 30-40 nm. In our case, the leads may act as gates and, consequently, the $pn$-interfaces near the contacts remain sharp, in agreement with theoretical estimates. Our interfacial width of $d \sim 20$ nm, on the other hand, is much smaller than found using scanning photocurrent microscopy on non-suspended samples fabricated on silicon dioxide \cite{Kern2008,Mueller2010}.

Our primary fit parameter $\Delta E_{\textrm{FM}}$ takes into account the standard electrostatics in the contact region. Our analysis indicates a clear success of electrostatic analysis, and the results verify the role of large contact capacitance arising due to charge transfer between the contact metal and graphene. This leads to pinning of Fermi level at the contact, which is consistent with findings in Refs. \onlinecite{Kern2008} and \onlinecite{Mueller2010}. Near the Dirac point, we find modifications from the standard electrostatic doping picture, which are presumably  related with the neglect of proper screening  treatment and  nonuniformities in the charge distribution near the Dirac point.

Finally, our analysis is based on a rather idealized theoretical model.  Additional effects can be included in to our theoretical
model, for example the broadening of the density of states in the graphene due to inhomogeneities and due to coupling to metal. Clearly, the inclusion of these effects would broaden the
conductance and the Fano factor characteristics, resulting in a better fit with our experimental data.  The disadvantage of this  phenomenological approach is that supplementary knowledge on additional parameters would be needed. We did check, however, what happens if a finite contact resistance is included using Eq. (\ref{contactresistance}). From the simulations, we find that a significant discrepancy in the overall value of the Fano factor starts to appear for $t_{c} < 0.9$: the Fano factor increases quickly over the entire gate voltage range and the fitting becomes difficult, no matter how large doping is used. This constrains the value of $t_{c}$ to $\sim 0.95 -1$, but the improvements in the fitting achieved within this range are negligibly small.

In conclusion, we have developed  a practical transport model for analyzing transport in ballistic, suspended graphene samples. Our analysis shows how the doping in graphene depends on the difference between the work function of the metal and graphene as well as on the applied gate voltage measured from the Dirac point which is given by the work function difference between the backgate material and graphene. By combining conductance and shot noise experiments performed  on a high-quality suspended graphene sample, we have determined all the relevant parameters which are involved in the electrostatics of the contact and  in the Klein tunneling of graphene.  When comparing with DFT calculations \cite{giovannetti2008}, we find a semiquantitative agreement for the graphene-modified metal work functions as well as for the distance between the charge separation layers which govern the contact capacitance between the metal and graphene. The small charge layer separation ($\sim 1$ {\AA}) leads to a large contact capacitance which is responsible for a rather weak tunability of the Fermi level position under the contact.

\section*{Acknowledgements}
We acknowledge fruitful discussions with D. Cox, V. Falko, T. Heikkil\"{a}, M. Katsnelson,  M.-H. Liu, and A. De Sanctis.  Our work was supported by the Academy of Finland (contract 250280, LTQ CoE) and by the Graphene Flagship project. This research project made use of the Aalto University Cryohall infrastructure. S.R. and M.F.C acknowledge financial support from EPSRC (Grant no. EP/J000396/1, EP/K017160/1, EP/K010050/1, EPG036101/1, EP/M001024/1, EPM002438/1) and from Royal Society international Exchanges Scheme 2012/R3 and 2013/R2.

\appendix

\section{The Fermi electric flux}
\label{appendix_xiF}

The Fermi electric flux
\begin{equation}
\zeta_{\rm F}= \frac{\sqrt{\pi}\hbar v_{\rm F}}{e}.
\end{equation}
 is related to the concept of quantum capacitance per unit area of graphene,
\begin{equation}
{\cal C}_{q}= e^2 {\cal D}(\Delta E_{\rm F}) = e^2 \frac{dn}{d(\Delta E_{\rm F})},
\end{equation}
where $n$ is the number of excess (negatively charged) carriers per unit area, ${\cal D}(\Delta E_{\rm F})$ is the density of states of graphene,
and $\Delta E_{\rm F}$ is the energy of the Fermi level measured from the Dirac point,
\begin{equation}
\Delta E_{\rm F} = \hbar v_{\rm F}{\rm sgn}\left[ n \right]\sqrt{\pi|n|} = e \zeta_{\rm F} {\rm sgn}[n]\sqrt{|n|}.
\end{equation}
Another useful formula is $|n|=(\Delta E_{\rm F}/e\zeta_{\rm F})^2$.
With these notations one immediately obtains
\begin{equation}
\Delta E_{\rm F} = \frac{1}{2} C_{q} \zeta_{\rm F}^2,
\end{equation}
resembling the formula for the energy $\frac{1}{2}C V^2$ of a capacitor
 $C$ charged by a fixed voltage $V$. Therefore the Fermi electric flux
is the electric flux through the plates of a capacitor of unit area and unit distance between the plates, with capacitance equal to the quantum capacitance and charged to an electrostatic energy $\Delta E_{\rm F}$.

The Fermi electric flux is a useful constant also for studying the inductive properties of graphene. The duality between the electric and magnetic properties is manifested in this case through the simple relation $\xi_{\rm F} = v_{\rm F} \Phi_{0}/\sqrt{\pi}$, where $\Phi_{0} = h/2e$ is the flux quantum. Then, the graphene kinetic inductance can be defined by regarding the Fermi energy $\epsilon_{\rm F}$ as a magnetic energy associated with the magnetic flux $\xi_{\rm F}/v_{\rm F}$, yielding \begin{equation} L_{\rm G,kin} = \frac{\xi_{\rm F}^{2}}{v_{\rm F}^2\epsilon_{\rm F}} \end{equation} in agreement with the known result for the kinetic inductance of graphene \cite{kinetic_inductance_philipkim}.

Another connection can be made with the fine structure constant,
$\alpha = e^2/4\pi \epsilon_{0}\hbar c\approx 1/137$, which has an essential role in determining the optical properties of graphene \cite{nair2008}.
We obtain
\begin{equation}
\zeta_{\rm F} = \frac{ev_{\rm F}}{4\sqrt{\pi}\epsilon_{0}\alpha c}.
\end{equation}
Similarly, one can introduce a graphene fine structure constant $\alpha_{\rm G} = e^2 / 4\pi \epsilon_{0} \hbar v_{\rm F} = \alpha c/v_{\rm F} \approx 2$, obtaining
$\zeta_{\rm F} = e /4\sqrt{\pi}\epsilon_{0}\alpha_{\rm G}$.

\section{Conductance for the whole barrier}
\label{incoherent}

Equation (\ref{totaltransmission})  is well known in mesoscopic physics and can be found in some textbooks, \emph{e.g.}  by S. Datta \cite{dattabook} and by Yu. Nazarov \& Ya. Blanter \cite{nazarov}. It is derived by summing over all the \FP reflection and transmission processes and neglecting the interference term in the final result \cite{dattabook}. Here we give an alternative, simpler proof, starting from the beginning with the assumption that the currents simply add up without any interference term.

\begin{figure}
\centering
\includegraphics[width=75mm]{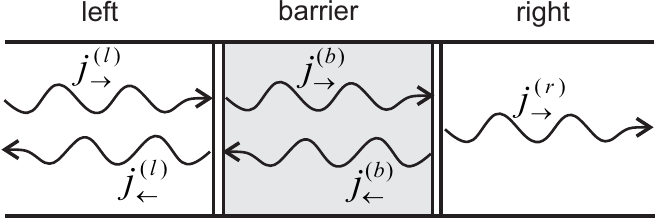}
\caption{(Color online) Generic schematic for tunneling through two interfaces, separating a middle region $(b)$ from the left $(l)$ and the right $(r)$ regions. }
\label{fig:incoh}
\end{figure}

 We consider the generic problem of incoherent tunneling through two interfaces, interface 1 separating the left region from a middle region and interface 2 separating the middle region from the right region, see Fig. \ref{fig:incoh}. Typically in this type of problems, the middle region is associated with a barrier, so we will denote it as such. The barrier has transmission coefficients $T_1$ on the left side and $T_{2}$ on the right side; the corresponding reflection coefficients are denoted by $R_{1}=1-T_{1}$ and $R_{2}=1-T_{2}$.

We assume that a wave producing a current $j^{(l)}_{\rightarrow}$ propagates from the left $(l)$ region towards the barrier. Part of this wave will get reflected back to the left into $j^{(l)}_{\leftarrow}$, and part of it gets transmitted in the region under the barrier $(b)$, with the current $j^{(b)}_{\rightarrow}$. At the right side of the barrier, the incoming current $j^{(b)}_{\rightarrow}$ gets transmitted to the region on the right of the barrier $(r)$ as $j^{(r)}_{\rightarrow}$ and part of it is reflected back under the barrier as $j^{(b)}_{\leftarrow}$. The wave $j^{(b)}_{\leftarrow}$ travels backwards (to the left) towards the left side of the barrier, where it is partly reflected and partly transmitted, contributing to $j^{(b)}_{\rightarrow}$ and $j^{(l)}_{\leftarrow}$ respectively. By using the fact that in the absence of interference the currents will just sum up, we get
\begin{eqnarray}
j^{(r)}_{\rightarrow} &=& T_{2}j^{(b)}_{\rightarrow},\label{unu}\\
j^{(b)}_{\leftarrow} &=& R_{2}j^{(b)}_{\rightarrow}, \label{doi}\\
j^{(b)}_{\rightarrow} &=& T_{1}j^{(l)}_{\rightarrow} + R_{1}j^{(b)}_{\leftarrow},\label{trei}\\
j^{(l)}_{\leftarrow} &=& R_{1}j^{(l)}_{\rightarrow} + T_{1}j^{(b)}_{\leftarrow} .\label{patru}
\end{eqnarray}
Adding the first pair of equations Eqs. (\ref{unu}-\ref{doi}) yields the conservation law for the current at the
$(b)-(r)$ interface, $j^{(r)}_{\rightarrow} =
j^{(b)}_{\rightarrow}- j^{(b)}_{\leftarrow} $, while adding the second pair Eqs. (\ref{trei}-\ref{patru}) gives the conservation law at the $(l)-(b)$ interface,
$ j^{(l)}_{\rightarrow}  - j^{(l)}_{\leftarrow} =
j^{(b)}_{\rightarrow}-j^{(b)}_{\leftarrow}$.
 From these two equations we get $j^{(l)}_{\rightarrow} -j^{(l)}_{\leftarrow} =j^{(r)}_{\rightarrow}$, which is a conservation law across the entire barrier. Finally, by combining the Eqs. (\ref{unu}-\ref{trei}) above we get
\begin{equation}
j^{(r)}_{\rightarrow} = \frac{T_{1}T_{2}}{1-R_{1}R_{2}}j^{(l)}_{\rightarrow},
\end{equation}
from which we can read directly the overall transmission coefficient across the barrier, $T_{\rm tot}= T_{1}T_{2}/(1-R_{1}R_{2})$. For $T_{1}=T_{2}=T$ this yields
\begin{equation}
T_{\rm tot}= \frac{1}{2T^{-1}-1}, \label{transmissionS}
\end{equation}
which was given in Eq. (\ref{totaltransmission}) of the main text.
Similarly, from Eqs. (\ref{doi}-\ref{patru}) we obtain
\begin{equation}
j^{(l)}_{\leftarrow} = \frac{R_{1} + R_{2} - 2R_{1}R_{2}}{1-R_{1}R_{2}}j^{(l)}_{\rightarrow},
\end{equation}
from which we get the overall reflection coefficient across the barrier.
$R_{\rm tot}= (R_{1} + R_{2} - 2R_{1}R_{2})/(1-R_{1}R_{2})$.

\section{Finite bias}
\label{finitebias}

The model presented in the paper does not include the small voltage bias used in the experiment to create a nonzero electric current between the left and the right contacts. A small finite bias voltage $V_b$ can be easily included, as detailed below.
We assume that this bias is equally distributed, $V_{b}/2$ between the left metal and the graphene and $-V_{b}/2$ between the right metal and the graphene sheet. We also assume that the currents are small enough so that the equilibrium structure of the energy levels in graphene is not considerably changed. The bias only provides a slightly higher Fermi level in the metal from which electrons are injected at the left side and holes are injected at the right side. Due to electron-hole symmetry, the total current and the total noise can be calculated by adding the contributions corresponding to electron and hole transport \cite{Sonin2008}.

The electron injected from the left into the graphene sheet will have a Fermi energy shift in the metal contact region $\Delta E_{\rm FM} + eV_{b}/2$, and a Fermi energy shift $ \Delta E_{\rm FG} + eV_{b}/2$ in the suspended region. Similarly, for the hole injected from the right side, we will have the shifts $\Delta E_{\rm FM} - eV_{b}/2$
and $\Delta E_{\rm FG} - eV_{b}/2$. The current at a bias voltage $V_{b}+dV_{b}$ can be calculated in this model by taking into account that this voltage will be distributed onto both junctions,
$dj=\sigma^{(+)}dV_{b}/2 + \sigma^{(-)}dV_{b}/2$.
The measured conductance for a bias voltage $V_{b}$ is defined as $\sigma =dj /dV_{b}$, yielding
\begin{equation}
\sigma =\frac{1}{2}\left[\sigma^{(+)}+\sigma^{(-)}\right].
\end{equation}
Similarly, the differential noise for a bias voltage $V_{b}$ is
\begin{equation}
s =\frac{1}{2}\left[s^{(+)}+s^{(-)}\right],
\end{equation}
and the Fano factor becomes
\begin{equation}
F = \frac{s}{\sigma }.
\end{equation}
In these equations, $\sigma^{(\pm )}$ and
$s^{(\pm )}$ are the same as $\sigma$ and $s$ introduced in Sec. \ref{cond}, but are now calculated for electrons with energies $\Delta E_{\rm FM}\pm eV_{b}/2$, $\Delta E_{\rm FG}\pm eV_{b}/2$ shifted from the equilibrium values calculated in Eqs. (\ref{sol1}) and (\ref{sol2}), and correspondingly shifted momenta  (see Eqs. (\ref{kFG}) and ( \ref{kFM})),
$k_{\rm FG}^{(\pm )} = \left|\Delta E_{\rm FM} \pm eV_{b}/2 \right|\hbar v_{\rm F}$ and
$k_{\rm FM}^{(\pm )} = \left|k_{\rm FM} \pm eV_{b}/2 \right|\hbar v_{\rm F}$.
To summarize,
\begin{eqnarray}
\sigma^{(\pm )}  &=& \frac{4e^{2}W}{\pi h}\int_{0}^{{\rm min}\left\{k_{\rm FM}^{(\pm)}, k_{\rm FG}^{(\pm )}\right\}}
d k_{y} T_{\rm tot}^{(\pm )}, \\
 s^{(\pm )} &=& \frac{4e^{2}W}{\pi h}\int_{0}^{{\rm min}\left\{k_{\rm FM}^{(\pm )}, k_{\rm FG}^{(\pm )}\right\}}
d k_{y} \left[ 1- T_{\rm tot}^{(\pm )}\right]T_{\rm tot}^{(\pm )},
\end{eqnarray}
where  $T_{\rm tot}^{(\pm )}$ is defined according to Eq. (\ref{totaltransmission}), but at energies $\Delta E_{\rm FM}\pm eV_{b}/2$, $\Delta E_{\rm FG}\pm eV_{b}/2$ and at shifted momenta $k_{\rm FM}^{(\pm )}$, $k_{\rm FG}^{(\pm )}$.

Note that due to the assumptions above, the impact parameter $p_{0}$ and the slope $a$ will not change, since these are calculated from the equilibrium values $eV_{0}= \Delta E_{\rm FM}- \Delta E_{\rm FG}$, and $a=\sqrt{\pi}\vert V_{0}\vert/\zeta_{\rm F}d$. Using this modified model we can account for the effect of a small bias voltage. The simulation show that, as expected, that this produces a slight broadening of the ideal (zero-bias) conductance and Fano factor, and does not have a significant effect for the determination of the doping parameters.

\bibliography{Collection}

\begin{thebibliography}{58}%
\makeatletter
\providecommand \@ifxundefined [1]{%
 \@ifx{#1\undefined}
}%
\providecommand \@ifnum [1]{%
 \ifnum #1\expandafter \@firstoftwo
 \else \expandafter \@secondoftwo
 \fi
}%
\providecommand \@ifx [1]{%
 \ifx #1\expandafter \@firstoftwo
 \else \expandafter \@secondoftwo
 \fi
}%
\providecommand \natexlab [1]{#1}%
\providecommand \enquote  [1]{``#1''}%
\providecommand \bibnamefont  [1]{#1}%
\providecommand \bibfnamefont [1]{#1}%
\providecommand \citenamefont [1]{#1}%
\providecommand \href@noop [0]{\@secondoftwo}%
\providecommand \href [0]{\begingroup \@sanitize@url \@href}%
\providecommand \@href[1]{\@@startlink{#1}\@@href}%
\providecommand \@@href[1]{\endgroup#1\@@endlink}%
\providecommand \@sanitize@url [0]{\catcode `\\12\catcode `\$12\catcode
  `\&12\catcode `\#12\catcode `\^12\catcode `\_12\catcode `\%12\relax}%
\providecommand \@@startlink[1]{}%
\providecommand \@@endlink[0]{}%
\providecommand \url  [0]{\begingroup\@sanitize@url \@url }%
\providecommand \@url [1]{\endgroup\@href {#1}{\urlprefix }}%
\providecommand \urlprefix  [0]{URL }%
\providecommand \Eprint [0]{\href }%
\providecommand \doibase [0]{http://dx.doi.org/}%
\providecommand \selectlanguage [0]{\@gobble}%
\providecommand \bibinfo  [0]{\@secondoftwo}%
\providecommand \bibfield  [0]{\@secondoftwo}%
\providecommand \translation [1]{[#1]}%
\providecommand \BibitemOpen [0]{}%
\providecommand \bibitemStop [0]{}%
\providecommand \bibitemNoStop [0]{.\EOS\space}%
\providecommand \EOS [0]{\spacefactor3000\relax}%
\providecommand \BibitemShut  [1]{\csname bibitem#1\endcsname}%
\let\auto@bib@innerbib\@empty
\bibitem [{\citenamefont {Tworzydlo}\ \emph {et~al.}(2006)\citenamefont
  {Tworzydlo}, \citenamefont {Trauzettel}, \citenamefont {Titov}, \citenamefont
  {Rycerz},\ and\ \citenamefont {Beenakker}}]{Tworzydio2006}%
  \BibitemOpen
  \bibfield  {author} {\bibinfo {author} {\bibfnamefont {J.}~\bibnamefont
  {Tworzydlo}}, \bibinfo {author} {\bibfnamefont {B.}~\bibnamefont
  {Trauzettel}}, \bibinfo {author} {\bibfnamefont {M.}~\bibnamefont {Titov}},
  \bibinfo {author} {\bibfnamefont {A.}~\bibnamefont {Rycerz}}, \ and\ \bibinfo
  {author} {\bibfnamefont {C.}~\bibnamefont {Beenakker}},\ }\href {\doibase
  10.1103/PhysRevLett.96.246802} {\bibfield  {journal} {\bibinfo  {journal}
  {Phys. Rev. Lett.}\ }\textbf {\bibinfo {volume} {96}},\ \bibinfo {pages}
  {246802} (\bibinfo {year} {2006})}\BibitemShut {NoStop}%
\bibitem [{\citenamefont {Katsnelson}\ \emph {et~al.}(2006)\citenamefont
  {Katsnelson}, \citenamefont {Novoselov},\ and\ \citenamefont
  {Geim}}]{Katsnelson2006b}%
  \BibitemOpen
  \bibfield  {author} {\bibinfo {author} {\bibfnamefont {M.~I.}\ \bibnamefont
  {Katsnelson}}, \bibinfo {author} {\bibfnamefont {K.~S.}\ \bibnamefont
  {Novoselov}}, \ and\ \bibinfo {author} {\bibfnamefont {A.~K.}\ \bibnamefont
  {Geim}},\ }\href@noop {} {\bibfield  {journal} {\bibinfo  {journal} {Nat.
  Phys.}\ }\textbf {\bibinfo {volume} {2}},\ \bibinfo {pages} {620} (\bibinfo
  {year} {2006})}\BibitemShut {NoStop}%
\bibitem [{\citenamefont {Katsnelson}(2006)}]{katsnelson2006a}%
  \BibitemOpen
  \bibfield  {author} {\bibinfo {author} {\bibfnamefont {M.~I.}\ \bibnamefont
  {Katsnelson}},\ }\href@noop {} {\bibfield  {journal} {\bibinfo  {journal}
  {Eur. Phys. J. B}\ }\textbf {\bibinfo {volume} {51}},\ \bibinfo {pages} {157}
  (\bibinfo {year} {2006})}\BibitemShut {NoStop}%
\bibitem [{\citenamefont {Miao}\ \emph {et~al.}(2007)\citenamefont {Miao},
  \citenamefont {Wijeratne}, \citenamefont {Zhang}, \citenamefont {Coskun},
  \citenamefont {Bao},\ and\ \citenamefont {Lau}}]{Miao2007a}%
  \BibitemOpen
  \bibfield  {author} {\bibinfo {author} {\bibfnamefont {F.}~\bibnamefont
  {Miao}}, \bibinfo {author} {\bibfnamefont {S.}~\bibnamefont {Wijeratne}},
  \bibinfo {author} {\bibfnamefont {Y.}~\bibnamefont {Zhang}}, \bibinfo
  {author} {\bibfnamefont {U.~C.}\ \bibnamefont {Coskun}}, \bibinfo {author}
  {\bibfnamefont {W.}~\bibnamefont {Bao}}, \ and\ \bibinfo {author}
  {\bibfnamefont {C.~N.}\ \bibnamefont {Lau}},\ }\href {\doibase
  10.1126/science.1144359} {\bibfield  {journal} {\bibinfo  {journal}
  {Science}\ }\textbf {\bibinfo {volume} {317}},\ \bibinfo {pages} {1530}
  (\bibinfo {year} {2007})}\BibitemShut {NoStop}%
\bibitem [{\citenamefont {Danneau}\ \emph
  {et~al.}(2008{\natexlab{a}})\citenamefont {Danneau}, \citenamefont {Wu},
  \citenamefont {Craciun}, \citenamefont {Russo}, \citenamefont {Tomi},
  \citenamefont {Salmilehto}, \citenamefont {Morpurgo},\ and\ \citenamefont
  {Hakonen}}]{Danneau2008}%
  \BibitemOpen
  \bibfield  {author} {\bibinfo {author} {\bibfnamefont {R.}~\bibnamefont
  {Danneau}}, \bibinfo {author} {\bibfnamefont {F.}~\bibnamefont {Wu}},
  \bibinfo {author} {\bibfnamefont {M.}~\bibnamefont {Craciun}}, \bibinfo
  {author} {\bibfnamefont {S.}~\bibnamefont {Russo}}, \bibinfo {author}
  {\bibfnamefont {M.}~\bibnamefont {Tomi}}, \bibinfo {author} {\bibfnamefont
  {J.}~\bibnamefont {Salmilehto}}, \bibinfo {author} {\bibfnamefont
  {A.}~\bibnamefont {Morpurgo}}, \ and\ \bibinfo {author} {\bibfnamefont
  {P.}~\bibnamefont {Hakonen}},\ }\href {\doibase
  10.1103/PhysRevLett.100.196802} {\bibfield  {journal} {\bibinfo  {journal}
  {Phys. Rev. Lett.}\ }\textbf {\bibinfo {volume} {100}},\ \bibinfo {pages}
  {196802} (\bibinfo {year} {2008}{\natexlab{a}})}\BibitemShut {NoStop}%
\bibitem [{\citenamefont {Du}\ \emph {et~al.}(2008)\citenamefont {Du},
  \citenamefont {Skachko}, \citenamefont {Barker},\ and\ \citenamefont
  {Andrei}}]{Du2008}%
  \BibitemOpen
  \bibfield  {author} {\bibinfo {author} {\bibfnamefont {X.}~\bibnamefont
  {Du}}, \bibinfo {author} {\bibfnamefont {I.}~\bibnamefont {Skachko}},
  \bibinfo {author} {\bibfnamefont {A.}~\bibnamefont {Barker}}, \ and\ \bibinfo
  {author} {\bibfnamefont {E.~Y.}\ \bibnamefont {Andrei}},\ }\href@noop {}
  {\bibfield  {journal} {\bibinfo  {journal} {Nat. Nanotechnol.}\ }\textbf
  {\bibinfo {volume} {3}},\ \bibinfo {pages} {491} (\bibinfo {year}
  {2008})}\BibitemShut {NoStop}%
\bibitem [{\citenamefont {Sonin}(2008)}]{Sonin2008}%
  \BibitemOpen
  \bibfield  {author} {\bibinfo {author} {\bibfnamefont {E.~B.}\ \bibnamefont
  {Sonin}},\ }\href {\doibase 10.1103/PhysRevB.77.233408} {\bibfield  {journal}
  {\bibinfo  {journal} {Phys. Rev. B}\ }\textbf {\bibinfo {volume} {77}},\
  \bibinfo {pages} {233408} (\bibinfo {year} {2008})}\BibitemShut {NoStop}%
\bibitem [{\citenamefont {Cheianov}\ and\ \citenamefont
  {Fal'ko}(2006)}]{Cheianov2006}%
  \BibitemOpen
  \bibfield  {author} {\bibinfo {author} {\bibfnamefont {V.~V.}\ \bibnamefont
  {Cheianov}}\ and\ \bibinfo {author} {\bibfnamefont {V.~I.}\ \bibnamefont
  {Fal'ko}},\ }\href {\doibase 10.1103/PhysRevB.74.041403} {\bibfield
  {journal} {\bibinfo  {journal} {Phys. Rev. B}\ }\textbf {\bibinfo {volume}
  {74}},\ \bibinfo {pages} {041403(R)} (\bibinfo {year} {2006})}\BibitemShut
  {NoStop}%
\bibitem [{\citenamefont {Shytov}\ \emph {et~al.}(2008)\citenamefont {Shytov},
  \citenamefont {Rudner},\ and\ \citenamefont {Levitov}}]{Shytov2008}%
  \BibitemOpen
  \bibfield  {author} {\bibinfo {author} {\bibfnamefont {A.}~\bibnamefont
  {Shytov}}, \bibinfo {author} {\bibfnamefont {M.}~\bibnamefont {Rudner}}, \
  and\ \bibinfo {author} {\bibfnamefont {L.}~\bibnamefont {Levitov}},\ }\href
  {\doibase 10.1103/PhysRevLett.101.156804} {\bibfield  {journal} {\bibinfo
  {journal} {Phys. Rev. Lett.}\ }\textbf {\bibinfo {volume} {101}},\ \bibinfo
  {pages} {156804} (\bibinfo {year} {2008})}\BibitemShut {NoStop}%
\bibitem [{\citenamefont {Cayssol}\ \emph {et~al.}(2009)\citenamefont
  {Cayssol}, \citenamefont {Huard},\ and\ \citenamefont
  {Goldhaber-Gordon}}]{Cayssol2009}%
  \BibitemOpen
  \bibfield  {author} {\bibinfo {author} {\bibfnamefont {J.}~\bibnamefont
  {Cayssol}}, \bibinfo {author} {\bibfnamefont {B.}~\bibnamefont {Huard}}, \
  and\ \bibinfo {author} {\bibfnamefont {D.}~\bibnamefont {Goldhaber-Gordon}},\
  }\href {\doibase 10.1103/PhysRevB.79.075428} {\bibfield  {journal} {\bibinfo
  {journal} {Phys. Rev. B}\ }\textbf {\bibinfo {volume} {79}},\ \bibinfo
  {pages} {075428} (\bibinfo {year} {2009})},\ \Eprint
  {http://arxiv.org/abs/arXiv:0810.4568v2} {arXiv:arXiv:0810.4568v2}
  \BibitemShut {NoStop}%
\bibitem [{\citenamefont {Huard}\ \emph {et~al.}(2007)\citenamefont {Huard},
  \citenamefont {Sulpizio}, \citenamefont {Stander}, \citenamefont {Todd},
  \citenamefont {Yang},\ and\ \citenamefont {Goldhaber-Gordon}}]{Huard2007}%
  \BibitemOpen
  \bibfield  {author} {\bibinfo {author} {\bibfnamefont {B.}~\bibnamefont
  {Huard}}, \bibinfo {author} {\bibfnamefont {J.}~\bibnamefont {Sulpizio}},
  \bibinfo {author} {\bibfnamefont {N.}~\bibnamefont {Stander}}, \bibinfo
  {author} {\bibfnamefont {K.}~\bibnamefont {Todd}}, \bibinfo {author}
  {\bibfnamefont {B.}~\bibnamefont {Yang}}, \ and\ \bibinfo {author}
  {\bibfnamefont {D.}~\bibnamefont {Goldhaber-Gordon}},\ }\href {\doibase
  10.1103/PhysRevLett.98.236803} {\bibfield  {journal} {\bibinfo  {journal}
  {Phys. Rev. Lett.}\ }\textbf {\bibinfo {volume} {98}},\ \bibinfo {pages}
  {236803} (\bibinfo {year} {2007})}\BibitemShut {NoStop}%
\bibitem [{\citenamefont {Huard}\ \emph {et~al.}(2008)\citenamefont {Huard},
  \citenamefont {Stander}, \citenamefont {Sulpizio},\ and\ \citenamefont
  {Goldhaber-Gordon}}]{Huard2008}%
  \BibitemOpen
  \bibfield  {author} {\bibinfo {author} {\bibfnamefont {B.}~\bibnamefont
  {Huard}}, \bibinfo {author} {\bibfnamefont {N.}~\bibnamefont {Stander}},
  \bibinfo {author} {\bibfnamefont {J.}~\bibnamefont {Sulpizio}}, \ and\
  \bibinfo {author} {\bibfnamefont {D.}~\bibnamefont {Goldhaber-Gordon}},\
  }\href {\doibase 10.1103/PhysRevB.78.121402} {\bibfield  {journal} {\bibinfo
  {journal} {Phys. Rev. B}\ }\textbf {\bibinfo {volume} {78}},\ \bibinfo
  {pages} {121402} (\bibinfo {year} {2008})}\BibitemShut {NoStop}%
\bibitem [{\citenamefont {Stander}\ \emph {et~al.}(2009)\citenamefont
  {Stander}, \citenamefont {Huard},\ and\ \citenamefont
  {Goldhaber-Gordon}}]{Stander2009}%
  \BibitemOpen
  \bibfield  {author} {\bibinfo {author} {\bibfnamefont {N.}~\bibnamefont
  {Stander}}, \bibinfo {author} {\bibfnamefont {B.}~\bibnamefont {Huard}}, \
  and\ \bibinfo {author} {\bibfnamefont {D.}~\bibnamefont {Goldhaber-Gordon}},\
  }\href {\doibase 10.1103/PhysRevLett.102.026807} {\bibfield  {journal}
  {\bibinfo  {journal} {Phys. Rev. Lett.}\ }\textbf {\bibinfo {volume} {102}},\
  \bibinfo {pages} {026807} (\bibinfo {year} {2009})}\BibitemShut {NoStop}%
\bibitem [{\citenamefont {Rickhaus}\ \emph {et~al.}(2013)\citenamefont
  {Rickhaus}, \citenamefont {Maurand}, \citenamefont {Liu}, \citenamefont
  {Weiss}, \citenamefont {Richter},\ and\ \citenamefont
  {Sch\"{o}nenberger}}]{rickhaus2013}%
  \BibitemOpen
  \bibfield  {author} {\bibinfo {author} {\bibfnamefont {P.}~\bibnamefont
  {Rickhaus}}, \bibinfo {author} {\bibfnamefont {R.}~\bibnamefont {Maurand}},
  \bibinfo {author} {\bibfnamefont {M.-H.}\ \bibnamefont {Liu}}, \bibinfo
  {author} {\bibfnamefont {M.}~\bibnamefont {Weiss}}, \bibinfo {author}
  {\bibfnamefont {K.}~\bibnamefont {Richter}}, \ and\ \bibinfo {author}
  {\bibfnamefont {C.}~\bibnamefont {Sch\"{o}nenberger}},\ }\href {\doibase
  10.1038/ncomms3342} {\bibfield  {journal} {\bibinfo  {journal} {Nat.
  Commun.}\ }\textbf {\bibinfo {volume} {4}},\ \bibinfo {pages} {2342}
  (\bibinfo {year} {2013})}\BibitemShut {NoStop}%
\bibitem [{\citenamefont {Oksanen}\ \emph {et~al.}(2014)\citenamefont
  {Oksanen}, \citenamefont {Uppstu}, \citenamefont {Laitinen}, \citenamefont
  {Cox}, \citenamefont {Craciun}, \citenamefont {Russo}, \citenamefont
  {Harju},\ and\ \citenamefont {Hakonen}}]{Oksanen2014}%
  \BibitemOpen
  \bibfield  {author} {\bibinfo {author} {\bibfnamefont {M.}~\bibnamefont
  {Oksanen}}, \bibinfo {author} {\bibfnamefont {A.}~\bibnamefont {Uppstu}},
  \bibinfo {author} {\bibfnamefont {A.}~\bibnamefont {Laitinen}}, \bibinfo
  {author} {\bibfnamefont {D.~J.}\ \bibnamefont {Cox}}, \bibinfo {author}
  {\bibfnamefont {M.~F.}\ \bibnamefont {Craciun}}, \bibinfo {author}
  {\bibfnamefont {S.}~\bibnamefont {Russo}}, \bibinfo {author} {\bibfnamefont
  {A.}~\bibnamefont {Harju}}, \ and\ \bibinfo {author} {\bibfnamefont
  {P.}~\bibnamefont {Hakonen}},\ }\href {\doibase 10.1103/PhysRevB.89.121414}
  {\bibfield  {journal} {\bibinfo  {journal} {Phys. Rev. B}\ }\textbf {\bibinfo
  {volume} {89}},\ \bibinfo {pages} {121414} (\bibinfo {year}
  {2014})}\BibitemShut {NoStop}%
\bibitem [{\citenamefont {Novikov}(2007)}]{Novikov2007}%
  \BibitemOpen
  \bibfield  {author} {\bibinfo {author} {\bibfnamefont {D.~S.}\ \bibnamefont
  {Novikov}},\ }\href {\doibase 10.1063/1.2779107} {\bibfield  {journal}
  {\bibinfo  {journal} {Appl. Phys. Lett.}\ }\textbf {\bibinfo {volume} {91}},\
  \bibinfo {pages} {102102} (\bibinfo {year} {2007})}\BibitemShut {NoStop}%
\bibitem [{\citenamefont {Chen}\ \emph {et~al.}(2008)\citenamefont {Chen},
  \citenamefont {Jang}, \citenamefont {Adam}, \citenamefont {Fuhrer},
  \citenamefont {Williams},\ and\ \citenamefont {Ishigami}}]{Chen2008a}%
  \BibitemOpen
  \bibfield  {author} {\bibinfo {author} {\bibfnamefont {J.-H.}\ \bibnamefont
  {Chen}}, \bibinfo {author} {\bibfnamefont {C.}~\bibnamefont {Jang}}, \bibinfo
  {author} {\bibfnamefont {S.}~\bibnamefont {Adam}}, \bibinfo {author}
  {\bibfnamefont {M.~S.}\ \bibnamefont {Fuhrer}}, \bibinfo {author}
  {\bibfnamefont {E.~D.}\ \bibnamefont {Williams}}, \ and\ \bibinfo {author}
  {\bibfnamefont {M.}~\bibnamefont {Ishigami}},\ }\href {\doibase
  10.1038/nphys935} {\bibfield  {journal} {\bibinfo  {journal} {Nat. Phys.}\
  }\textbf {\bibinfo {volume} {4}},\ \bibinfo {pages} {377} (\bibinfo {year}
  {2008})}\BibitemShut {NoStop}%
\bibitem [{\citenamefont {Young}\ and\ \citenamefont {Kim}(2009)}]{Young2009}%
  \BibitemOpen
  \bibfield  {author} {\bibinfo {author} {\bibfnamefont {A.~F.}\ \bibnamefont
  {Young}}\ and\ \bibinfo {author} {\bibfnamefont {P.}~\bibnamefont {Kim}},\
  }\href {\doibase 10.1038/nphys1198} {\bibfield  {journal} {\bibinfo
  {journal} {Nat. Phys.}\ }\textbf {\bibinfo {volume} {5}},\ \bibinfo {pages}
  {222} (\bibinfo {year} {2009})}\BibitemShut {NoStop}%
\bibitem [{\citenamefont {Young}\ and\ \citenamefont {Kim}(2011)}]{Young2011}%
  \BibitemOpen
  \bibfield  {author} {\bibinfo {author} {\bibfnamefont {A.~F.}\ \bibnamefont
  {Young}}\ and\ \bibinfo {author} {\bibfnamefont {P.}~\bibnamefont {Kim}},\
  }\href {\doibase 10.1146/annurev-conmatphys-062910-140458} {\bibfield
  {journal} {\bibinfo  {journal} {Annu. Rev. Condens. Matter Phys.}\ }\textbf
  {\bibinfo {volume} {2}},\ \bibinfo {pages} {101} (\bibinfo {year}
  {2011})}\BibitemShut {NoStop}%
\bibitem [{\citenamefont {Liu}\ \emph {et~al.}(2008)\citenamefont {Liu},
  \citenamefont {Velasco}, \citenamefont {Bao},\ and\ \citenamefont
  {Lau}}]{LiuG2008}%
  \BibitemOpen
  \bibfield  {author} {\bibinfo {author} {\bibfnamefont {G.}~\bibnamefont
  {Liu}}, \bibinfo {author} {\bibfnamefont {J.}~\bibnamefont {Velasco}},
  \bibinfo {author} {\bibfnamefont {W.}~\bibnamefont {Bao}}, \ and\ \bibinfo
  {author} {\bibfnamefont {C.~N.}\ \bibnamefont {Lau}},\ }\href {\doibase
  10.1063/1.2928234} {\bibfield  {journal} {\bibinfo  {journal} {Appl. Phys.
  Lett.}\ }\textbf {\bibinfo {volume} {92}},\ \bibinfo {pages} {203103}
  (\bibinfo {year} {2008})}\BibitemShut {NoStop}%
\bibitem [{\citenamefont {Gorbachev}\ \emph {et~al.}(2008)\citenamefont
  {Gorbachev}, \citenamefont {Mayorov}, \citenamefont {Savchenko},
  \citenamefont {Horsell},\ and\ \citenamefont {Guinea}}]{Gorbachev2008}%
  \BibitemOpen
  \bibfield  {author} {\bibinfo {author} {\bibfnamefont {R.~V.}\ \bibnamefont
  {Gorbachev}}, \bibinfo {author} {\bibfnamefont {A.~S.}\ \bibnamefont
  {Mayorov}}, \bibinfo {author} {\bibfnamefont {A.~K.}\ \bibnamefont
  {Savchenko}}, \bibinfo {author} {\bibfnamefont {D.~W.}\ \bibnamefont
  {Horsell}}, \ and\ \bibinfo {author} {\bibfnamefont {F.}~\bibnamefont
  {Guinea}},\ }\href {\doibase 10.1021/nl801059v} {\bibfield  {journal}
  {\bibinfo  {journal} {Nano Lett.}\ }\textbf {\bibinfo {volume} {8}},\
  \bibinfo {pages} {1995} (\bibinfo {year} {2008})}\BibitemShut {NoStop}%
\bibitem [{\citenamefont {Ferrari}(2014)}]{ferrari2015}%
  \BibitemOpen
  \bibfield  {author} {\bibinfo {author} {\bibfnamefont {A.~C.}\ \bibnamefont
  {Ferrari}},\ }\href {\doibase 10.1039/C4NR01600A} {\bibfield  {journal}
  {\bibinfo  {journal} {Nanoscale}\ }\textbf {\bibinfo {volume} {7}},\ \bibinfo
  {pages} {4598} (\bibinfo {year} {2014})}\BibitemShut {NoStop}%
\bibitem [{\citenamefont {Tielrooij}\ \emph {et~al.}(2015)\citenamefont
  {Tielrooij}, \citenamefont {Piatkowski}, \citenamefont {Massicotte},
  \citenamefont {Woessner}, \citenamefont {Ma}, \citenamefont {Lee},
  \citenamefont {Myhro}, \citenamefont {Lau}, \citenamefont {Jarillo-Herrero},
  \citenamefont {van Hulst},\ and\ \citenamefont {Koppens}}]{tielrooij2015}%
  \BibitemOpen
  \bibfield  {author} {\bibinfo {author} {\bibfnamefont {K.~J.}\ \bibnamefont
  {Tielrooij}}, \bibinfo {author} {\bibfnamefont {L.}~\bibnamefont
  {Piatkowski}}, \bibinfo {author} {\bibfnamefont {M.}~\bibnamefont
  {Massicotte}}, \bibinfo {author} {\bibfnamefont {A.}~\bibnamefont
  {Woessner}}, \bibinfo {author} {\bibfnamefont {Q.}~\bibnamefont {Ma}},
  \bibinfo {author} {\bibfnamefont {Y.}~\bibnamefont {Lee}}, \bibinfo {author}
  {\bibfnamefont {K.~S.}\ \bibnamefont {Myhro}}, \bibinfo {author}
  {\bibfnamefont {C.~N.}\ \bibnamefont {Lau}}, \bibinfo {author} {\bibfnamefont
  {P.}~\bibnamefont {Jarillo-Herrero}}, \bibinfo {author} {\bibfnamefont
  {N.~F.}\ \bibnamefont {van Hulst}}, \ and\ \bibinfo {author} {\bibfnamefont
  {F.~H.~L.}\ \bibnamefont {Koppens}},\ }\href {\doibase 10.1038/nnano.2015.54}
  {\bibfield  {journal} {\bibinfo  {journal} {Nat. Nanotechnol.}\ }\textbf
  {\bibinfo {volume} {10}},\ \bibinfo {pages} {437} (\bibinfo {year}
  {2015})}\BibitemShut {NoStop}%
\bibitem [{\citenamefont {Giovannetti}\ \emph {et~al.}(2008)\citenamefont
  {Giovannetti}, \citenamefont {Khomyakov}, \citenamefont {Brocks},
  \citenamefont {Karpan}, \citenamefont {van~den Brink},\ and\ \citenamefont
  {Kelly}}]{giovannetti2008}%
  \BibitemOpen
  \bibfield  {author} {\bibinfo {author} {\bibfnamefont {G.}~\bibnamefont
  {Giovannetti}}, \bibinfo {author} {\bibfnamefont {P.}~\bibnamefont
  {Khomyakov}}, \bibinfo {author} {\bibfnamefont {G.}~\bibnamefont {Brocks}},
  \bibinfo {author} {\bibfnamefont {V.}~\bibnamefont {Karpan}}, \bibinfo
  {author} {\bibfnamefont {J.}~\bibnamefont {van~den Brink}}, \ and\ \bibinfo
  {author} {\bibfnamefont {P.}~\bibnamefont {Kelly}},\ }\href {\doibase
  10.1103/PhysRevLett.101.026803} {\bibfield  {journal} {\bibinfo  {journal}
  {Phys. Rev. Lett.}\ }\textbf {\bibinfo {volume} {101}},\ \bibinfo {pages}
  {026803} (\bibinfo {year} {2008})}\BibitemShut {NoStop}%
\bibitem [{\citenamefont {Sonin}(2009)}]{Sonin2009}%
  \BibitemOpen
  \bibfield  {author} {\bibinfo {author} {\bibfnamefont {E.}~\bibnamefont
  {Sonin}},\ }\href {\doibase 10.1103/PhysRevB.79.195438} {\bibfield  {journal}
  {\bibinfo  {journal} {Phys. Rev. B}\ }\textbf {\bibinfo {volume} {79}},\
  \bibinfo {pages} {195438} (\bibinfo {year} {2009})}\BibitemShut {NoStop}%
\bibitem [{\citenamefont {Xia}\ \emph {et~al.}(2011)\citenamefont {Xia},
  \citenamefont {Perebeinos}, \citenamefont {Lin}, \citenamefont {Wu},\ and\
  \citenamefont {Avouris}}]{Xia2011}%
  \BibitemOpen
  \bibfield  {author} {\bibinfo {author} {\bibfnamefont {F.}~\bibnamefont
  {Xia}}, \bibinfo {author} {\bibfnamefont {V.}~\bibnamefont {Perebeinos}},
  \bibinfo {author} {\bibfnamefont {Y.-m.}\ \bibnamefont {Lin}}, \bibinfo
  {author} {\bibfnamefont {Y.}~\bibnamefont {Wu}}, \ and\ \bibinfo {author}
  {\bibfnamefont {P.}~\bibnamefont {Avouris}},\ }\href {\doibase
  10.1038/nnano.2011.6} {\bibfield  {journal} {\bibinfo  {journal} {Nat.
  Nanotechnol.}\ }\textbf {\bibinfo {volume} {6}},\ \bibinfo {pages} {179}
  (\bibinfo {year} {2011})}\BibitemShut {NoStop}%
\bibitem [{\citenamefont {Fang}\ \emph {et~al.}(2007)\citenamefont {Fang},
  \citenamefont {Konar}, \citenamefont {Xing},\ and\ \citenamefont
  {Jena}}]{fang2007}%
  \BibitemOpen
  \bibfield  {author} {\bibinfo {author} {\bibfnamefont {T.}~\bibnamefont
  {Fang}}, \bibinfo {author} {\bibfnamefont {A.}~\bibnamefont {Konar}},
  \bibinfo {author} {\bibfnamefont {H.}~\bibnamefont {Xing}}, \ and\ \bibinfo
  {author} {\bibfnamefont {D.}~\bibnamefont {Jena}},\ }\href {\doibase
  10.1063/1.2776887} {\bibfield  {journal} {\bibinfo  {journal} {Appl. Phys.
  Lett.}\ }\textbf {\bibinfo {volume} {91}},\ \bibinfo {pages} {092109}
  (\bibinfo {year} {2007})}\BibitemShut {NoStop}%
\bibitem [{\citenamefont {Fogler}\ \emph {et~al.}(2008)\citenamefont {Fogler},
  \citenamefont {Novikov}, \citenamefont {Glazman},\ and\ \citenamefont
  {Shklovskii}}]{Fogler2008}%
  \BibitemOpen
  \bibfield  {author} {\bibinfo {author} {\bibfnamefont {M.~M.}\ \bibnamefont
  {Fogler}}, \bibinfo {author} {\bibfnamefont {D.~S.}\ \bibnamefont {Novikov}},
  \bibinfo {author} {\bibfnamefont {L.~I.}\ \bibnamefont {Glazman}}, \ and\
  \bibinfo {author} {\bibfnamefont {B.~I.}\ \bibnamefont {Shklovskii}},\ }\href
  {\doibase 10.1103/PhysRevB.77.075420} {\bibfield  {journal} {\bibinfo
  {journal} {Phys. Rev. B}\ }\textbf {\bibinfo {volume} {77}},\ \bibinfo
  {pages} {075420} (\bibinfo {year} {2008})}\BibitemShut {NoStop}%
\bibitem [{\citenamefont {{Das Sarma}}\ \emph {et~al.}(2012)\citenamefont {{Das
  Sarma}}, \citenamefont {Adam}, \citenamefont {Hwang},\ and\ \citenamefont
  {Rossi}}]{DasSarma2012}%
  \BibitemOpen
  \bibfield  {author} {\bibinfo {author} {\bibfnamefont {S.}~\bibnamefont {{Das
  Sarma}}}, \bibinfo {author} {\bibfnamefont {S.}~\bibnamefont {Adam}},
  \bibinfo {author} {\bibfnamefont {E.~H.}\ \bibnamefont {Hwang}}, \ and\
  \bibinfo {author} {\bibfnamefont {E.}~\bibnamefont {Rossi}},\ }\href
  {http://dx.doi.org/10.1103/RevModPhys.83.407} {\bibfield  {journal} {\bibinfo
   {journal} {Rev. Mod. Phys.}\ }\textbf {\bibinfo {volume} {83}},\ \bibinfo
  {pages} {407} (\bibinfo {year} {2012})}\BibitemShut {NoStop}%
\bibitem [{\citenamefont {Khodkov}\ \emph {et~al.}(2012)\citenamefont
  {Khodkov}, \citenamefont {Withers}, \citenamefont {{Christopher Hudson}},
  \citenamefont {{Felicia Craciun}},\ and\ \citenamefont
  {Russo}}]{Khodkov2012}%
  \BibitemOpen
  \bibfield  {author} {\bibinfo {author} {\bibfnamefont {T.}~\bibnamefont
  {Khodkov}}, \bibinfo {author} {\bibfnamefont {F.}~\bibnamefont {Withers}},
  \bibinfo {author} {\bibfnamefont {D.}~\bibnamefont {{Christopher Hudson}}},
  \bibinfo {author} {\bibfnamefont {M.}~\bibnamefont {{Felicia Craciun}}}, \
  and\ \bibinfo {author} {\bibfnamefont {S.}~\bibnamefont {Russo}},\ }\href
  {\doibase 10.1063/1.3675337} {\bibfield  {journal} {\bibinfo  {journal}
  {Appl. Phys. Lett.}\ }\textbf {\bibinfo {volume} {100}},\ \bibinfo {pages}
  {013114} (\bibinfo {year} {2012})}\BibitemShut {NoStop}%
\bibitem [{\citenamefont {Khodkov}\ \emph {et~al.}(2015)\citenamefont
  {Khodkov}, \citenamefont {Khrapach}, \citenamefont {Craciun},\ and\
  \citenamefont {Russo}}]{Khodkov2015}%
  \BibitemOpen
  \bibfield  {author} {\bibinfo {author} {\bibfnamefont {T.}~\bibnamefont
  {Khodkov}}, \bibinfo {author} {\bibfnamefont {I.}~\bibnamefont {Khrapach}},
  \bibinfo {author} {\bibfnamefont {M.~F.}\ \bibnamefont {Craciun}}, \ and\
  \bibinfo {author} {\bibfnamefont {S.}~\bibnamefont {Russo}},\ }\href
  {\doibase 10.1021/acs.nanolett.5b00772} {\bibfield  {journal} {\bibinfo
  {journal} {Nano Lett.}\ }\textbf {\bibinfo {volume} {15}},\ \bibinfo {pages}
  {4429} (\bibinfo {year} {2015})}\BibitemShut {NoStop}%
\bibitem [{\citenamefont {Roschier}\ and\ \citenamefont
  {Hakonen}(2004)}]{Roschier2004}%
  \BibitemOpen
  \bibfield  {author} {\bibinfo {author} {\bibfnamefont {L.}~\bibnamefont
  {Roschier}}\ and\ \bibinfo {author} {\bibfnamefont {P.}~\bibnamefont
  {Hakonen}},\ }\href {\doibase 10.1016/j.cryogenics.2004.04.006} {\bibfield
  {journal} {\bibinfo  {journal} {Cryogenics (Guildf).}\ }\textbf {\bibinfo
  {volume} {44}},\ \bibinfo {pages} {783} (\bibinfo {year} {2004})}\BibitemShut
  {NoStop}%
\bibitem [{\citenamefont {Danneau}\ \emph
  {et~al.}(2008{\natexlab{b}})\citenamefont {Danneau}, \citenamefont {Wu},
  \citenamefont {Craciun}, \citenamefont {Russo}, \citenamefont {Tomi},
  \citenamefont {Salmilehto}, \citenamefont {Morpurgo},\ and\ \citenamefont
  {Hakonen}}]{Danneau2008b}%
  \BibitemOpen
  \bibfield  {author} {\bibinfo {author} {\bibfnamefont {R.}~\bibnamefont
  {Danneau}}, \bibinfo {author} {\bibfnamefont {F.}~\bibnamefont {Wu}},
  \bibinfo {author} {\bibfnamefont {M.~F.}\ \bibnamefont {Craciun}}, \bibinfo
  {author} {\bibfnamefont {S.}~\bibnamefont {Russo}}, \bibinfo {author}
  {\bibfnamefont {M.~Y.}\ \bibnamefont {Tomi}}, \bibinfo {author}
  {\bibfnamefont {J.}~\bibnamefont {Salmilehto}}, \bibinfo {author}
  {\bibfnamefont {A.~F.}\ \bibnamefont {Morpurgo}}, \ and\ \bibinfo {author}
  {\bibfnamefont {P.~J.}\ \bibnamefont {Hakonen}},\ }\href {\doibase
  10.1007/s10909-008-9837-z} {\bibfield  {journal} {\bibinfo  {journal} {J. Low
  Temp. Phys.}\ }\textbf {\bibinfo {volume} {153}},\ \bibinfo {pages} {374}
  (\bibinfo {year} {2008}{\natexlab{b}})}\BibitemShut {NoStop}%
\bibitem [{\citenamefont {Novikov}(2010)}]{novikov2010}%
  \BibitemOpen
  \bibfield  {author} {\bibinfo {author} {\bibfnamefont {A.}~\bibnamefont
  {Novikov}},\ }\href {\doibase 10.1016/j.sse.2009.09.005} {\bibfield
  {journal} {\bibinfo  {journal} {Solid. State. Electron.}\ }\textbf {\bibinfo
  {volume} {54}},\ \bibinfo {pages} {8} (\bibinfo {year} {2010})}\BibitemShut
  {NoStop}%
\bibitem [{\citenamefont {Oshima}\ and\ \citenamefont
  {Nagashima}(1997)}]{nagashio1997}%
  \BibitemOpen
  \bibfield  {author} {\bibinfo {author} {\bibfnamefont {C.}~\bibnamefont
  {Oshima}}\ and\ \bibinfo {author} {\bibfnamefont {A.}~\bibnamefont
  {Nagashima}},\ }\href {\doibase 10.1088/0953-8984/9/1/004} {\bibfield
  {journal} {\bibinfo  {journal} {J. Phys. Condens. Matter}\ }\textbf {\bibinfo
  {volume} {9}},\ \bibinfo {pages} {1} (\bibinfo {year} {1997})}\BibitemShut
  {NoStop}%
\bibitem [{\citenamefont {Zhang}\ and\ \citenamefont
  {Fogler}(2008)}]{Zhang2008}%
  \BibitemOpen
  \bibfield  {author} {\bibinfo {author} {\bibfnamefont {L.~M.}\ \bibnamefont
  {Zhang}}\ and\ \bibinfo {author} {\bibfnamefont {M.~M.}\ \bibnamefont
  {Fogler}},\ }\href {\doibase 10.1103/PhysRevLett.100.116804} {\bibfield
  {journal} {\bibinfo  {journal} {Phys. Rev. Lett.}\ }\textbf {\bibinfo
  {volume} {100}},\ \bibinfo {pages} {116804} (\bibinfo {year}
  {2008})}\BibitemShut {NoStop}%
\bibitem [{\citenamefont {Khomyakov}\ \emph {et~al.}(2010)\citenamefont
  {Khomyakov}, \citenamefont {Starikov}, \citenamefont {Brocks},\ and\
  \citenamefont {Kelly}}]{Khomyakov2010}%
  \BibitemOpen
  \bibfield  {author} {\bibinfo {author} {\bibfnamefont {P.~A.}\ \bibnamefont
  {Khomyakov}}, \bibinfo {author} {\bibfnamefont {A.~A.}\ \bibnamefont
  {Starikov}}, \bibinfo {author} {\bibfnamefont {G.}~\bibnamefont {Brocks}}, \
  and\ \bibinfo {author} {\bibfnamefont {P.~J.}\ \bibnamefont {Kelly}},\ }\href
  {\doibase 10.1103/PhysRevB.82.115437} {\bibfield  {journal} {\bibinfo
  {journal} {Phys. Rev. B}\ }\textbf {\bibinfo {volume} {82}},\ \bibinfo
  {pages} {115437} (\bibinfo {year} {2010})}\BibitemShut {NoStop}%
\bibitem [{\citenamefont {Voutilainen}\ \emph {et~al.}(2011)\citenamefont
  {Voutilainen}, \citenamefont {Fay}, \citenamefont {Hakkinen}, \citenamefont
  {Viljas}, \citenamefont {Heikkila},\ and\ \citenamefont
  {Hakonen}}]{Voutilainen2011}%
  \BibitemOpen
  \bibfield  {author} {\bibinfo {author} {\bibfnamefont {J.}~\bibnamefont
  {Voutilainen}}, \bibinfo {author} {\bibfnamefont {A.}~\bibnamefont {Fay}},
  \bibinfo {author} {\bibfnamefont {P.}~\bibnamefont {H{\"a}kkinen}}, \bibinfo
  {author} {\bibfnamefont {J.~K.}\ \bibnamefont {Viljas}}, \bibinfo {author}
  {\bibfnamefont {T.~T.}\ \bibnamefont {Heikkil{\"a}}}, \ and\ \bibinfo {author}
  {\bibfnamefont {P.~J.}\ \bibnamefont {Hakonen}},\ }\href {\doibase
  10.1103/PhysRevB.84.045419} {\bibfield  {journal} {\bibinfo  {journal} {Phys.
  Rev. B}\ }\textbf {\bibinfo {volume} {84}},\ \bibinfo {pages} {045419}
  (\bibinfo {year} {2011})}\BibitemShut {NoStop}%
\bibitem [{\citenamefont {Khomyakov}\ \emph {et~al.}(2009)\citenamefont
  {Khomyakov}, \citenamefont {Giovannetti}, \citenamefont {Rusu}, \citenamefont
  {Brocks}, \citenamefont {van~den Brink},\ and\ \citenamefont
  {Kelly}}]{Khomyakov2009}%
  \BibitemOpen
  \bibfield  {author} {\bibinfo {author} {\bibfnamefont {P.~A.}\ \bibnamefont
  {Khomyakov}}, \bibinfo {author} {\bibfnamefont {G.}~\bibnamefont
  {Giovannetti}}, \bibinfo {author} {\bibfnamefont {P.~C.}\ \bibnamefont
  {Rusu}}, \bibinfo {author} {\bibfnamefont {G.}~\bibnamefont {Brocks}},
  \bibinfo {author} {\bibfnamefont {J.}~\bibnamefont {van~den Brink}}, \ and\
  \bibinfo {author} {\bibfnamefont {P.~J.}\ \bibnamefont {Kelly}},\ }\href
  {\doibase 10.1103/PhysRevB.79.195425} {\bibfield  {journal} {\bibinfo
  {journal} {Phys. Rev. B}\ }\textbf {\bibinfo {volume} {79}},\ \bibinfo
  {pages} {195425} (\bibinfo {year} {2009})}\BibitemShut {NoStop}%
\bibitem [{\citenamefont {Malec}\ and\ \citenamefont
  {Davidovi\'{c}}(2011)}]{Malec2011}%
  \BibitemOpen
  \bibfield  {author} {\bibinfo {author} {\bibfnamefont {C.~E.}\ \bibnamefont
  {Malec}}\ and\ \bibinfo {author} {\bibfnamefont {D.}~\bibnamefont
  {Davidovi\'{c}}},\ }\href {\doibase 10.1103/PhysRevB.84.033407} {\bibfield
  {journal} {\bibinfo  {journal} {Phys. Rev. B}\ }\textbf {\bibinfo {volume}
  {84}},\ \bibinfo {pages} {033407} (\bibinfo {year} {2011})}\BibitemShut
  {NoStop}%
\bibitem [{\citenamefont {Song}\ \emph {et~al.}(2012)\citenamefont {Song},
  \citenamefont {Park}, \citenamefont {Sul},\ and\ \citenamefont
  {Cho}}]{Song2012c}%
  \BibitemOpen
  \bibfield  {author} {\bibinfo {author} {\bibfnamefont {S.~M.}\ \bibnamefont
  {Song}}, \bibinfo {author} {\bibfnamefont {J.~K.}\ \bibnamefont {Park}},
  \bibinfo {author} {\bibfnamefont {O.~J.}\ \bibnamefont {Sul}}, \ and\
  \bibinfo {author} {\bibfnamefont {B.~J.}\ \bibnamefont {Cho}},\ }\href
  {\doibase 10.1021/nl300266p} {\bibfield  {journal} {\bibinfo  {journal} {Nano
  Lett.}\ }\textbf {\bibinfo {volume} {12}},\ \bibinfo {pages} {3887} (\bibinfo
  {year} {2012})}\BibitemShut {NoStop}%
\bibitem [{\citenamefont {Nagashio}\ \emph {et~al.}(2009)\citenamefont
  {Nagashio}, \citenamefont {Nishimura}, \citenamefont {Kita},\ and\
  \citenamefont {Toriumi}}]{Nagashio2009}%
  \BibitemOpen
  \bibfield  {author} {\bibinfo {author} {\bibfnamefont {K.}~\bibnamefont
  {Nagashio}}, \bibinfo {author} {\bibfnamefont {T.}~\bibnamefont {Nishimura}},
  \bibinfo {author} {\bibfnamefont {K.}~\bibnamefont {Kita}}, \ and\ \bibinfo
  {author} {\bibfnamefont {A.}~\bibnamefont {Toriumi}},\ }\href {\doibase
  10.1109/IEDM.2009.5424297} {\bibfield  {journal} {\bibinfo  {journal} {2009
  IEEE Int. Electron Devices Meet.}\ }\textbf {\bibinfo {volume} {23}},\
  \bibinfo {pages} {2.1} (\bibinfo {year} {2009})}\BibitemShut {NoStop}%
\bibitem [{\citenamefont {Zan}\ \emph {et~al.}(2011)\citenamefont {Zan},
  \citenamefont {Bangert}, \citenamefont {Ramasse},\ and\ \citenamefont
  {Novoselov}}]{zan2011}%
  \BibitemOpen
  \bibfield  {author} {\bibinfo {author} {\bibfnamefont {R.}~\bibnamefont
  {Zan}}, \bibinfo {author} {\bibfnamefont {U.}~\bibnamefont {Bangert}},
  \bibinfo {author} {\bibfnamefont {Q.}~\bibnamefont {Ramasse}}, \ and\
  \bibinfo {author} {\bibfnamefont {K.~S.}\ \bibnamefont {Novoselov}},\ }\href
  {\doibase 10.1021/nl103980h} {\bibfield  {journal} {\bibinfo  {journal} {Nano
  Lett.}\ }\textbf {\bibinfo {volume} {11}},\ \bibinfo {pages} {1087} (\bibinfo
  {year} {2011})}\BibitemShut {NoStop}%
\bibitem [{\citenamefont {Ramasse}\ \emph {et~al.}(2012)\citenamefont
  {Ramasse}, \citenamefont {Zan}, \citenamefont {Bangert}, \citenamefont
  {Boukhvalov}, \citenamefont {Son},\ and\ \citenamefont
  {Novoselov}}]{ramasse2012}%
  \BibitemOpen
  \bibfield  {author} {\bibinfo {author} {\bibfnamefont {Q.~M.}\ \bibnamefont
  {Ramasse}}, \bibinfo {author} {\bibfnamefont {R.}~\bibnamefont {Zan}},
  \bibinfo {author} {\bibfnamefont {U.}~\bibnamefont {Bangert}}, \bibinfo
  {author} {\bibfnamefont {D.~W.}\ \bibnamefont {Boukhvalov}}, \bibinfo
  {author} {\bibfnamefont {Y.-W.}\ \bibnamefont {Son}}, \ and\ \bibinfo
  {author} {\bibfnamefont {K.~S.}\ \bibnamefont {Novoselov}},\ }\href {\doibase
  10.1021/nn300452y} {\bibfield  {journal} {\bibinfo  {journal} {ACS Nano}\
  }\textbf {\bibinfo {volume} {6}},\ \bibinfo {pages} {4063} (\bibinfo {year}
  {2012})}\BibitemShut {NoStop}%
\bibitem [{\citenamefont {Gong}\ \emph {et~al.}(2014)\citenamefont {Gong},
  \citenamefont {Mcdonnell}, \citenamefont {Qin}, \citenamefont {Azcatl},
  \citenamefont {Dong}, \citenamefont {Chabal}, \citenamefont {Cho},\ and\
  \citenamefont {Wallace}}]{gong2014}%
  \BibitemOpen
  \bibfield  {author} {\bibinfo {author} {\bibfnamefont {C.}~\bibnamefont
  {Gong}}, \bibinfo {author} {\bibfnamefont {S.}~\bibnamefont {Mcdonnell}},
  \bibinfo {author} {\bibfnamefont {X.}~\bibnamefont {Qin}}, \bibinfo {author}
  {\bibfnamefont {A.}~\bibnamefont {Azcatl}}, \bibinfo {author} {\bibfnamefont
  {H.}~\bibnamefont {Dong}}, \bibinfo {author} {\bibfnamefont {Y.~J.}\
  \bibnamefont {Chabal}}, \bibinfo {author} {\bibfnamefont {K.}~\bibnamefont
  {Cho}}, \ and\ \bibinfo {author} {\bibfnamefont {R.~M.}\ \bibnamefont
  {Wallace}},\ }\href@noop {} {\ ,\ \bibinfo {pages} {642} (\bibinfo {year}
  {2014})}\BibitemShut {NoStop}%
\bibitem [{\citenamefont {Shen}\ \emph {et~al.}(2013)\citenamefont {Shen},
  \citenamefont {Wang},\ and\ \citenamefont {Yu}}]{shen2013}%
  \BibitemOpen
  \bibfield  {author} {\bibinfo {author} {\bibfnamefont {X.}~\bibnamefont
  {Shen}}, \bibinfo {author} {\bibfnamefont {H.}~\bibnamefont {Wang}}, \ and\
  \bibinfo {author} {\bibfnamefont {T.}~\bibnamefont {Yu}},\ }\href {\doibase
  10.1039/c3nr33460k} {\bibfield  {journal} {\bibinfo  {journal} {Nanoscale}\
  }\textbf {\bibinfo {volume} {5}},\ \bibinfo {pages} {3352} (\bibinfo {year}
  {2013})}\BibitemShut {NoStop}%
\bibitem [{\citenamefont {Leong}\ \emph {et~al.}(2014)\citenamefont {Leong},
  \citenamefont {{Chang Tai Nai}},\ and\ \citenamefont {Thong}}]{leong2014}%
  \BibitemOpen
  \bibfield  {author} {\bibinfo {author} {\bibfnamefont {W.~S.}\ \bibnamefont
  {Leong}}, \bibinfo {author} {\bibnamefont {{Chang Tai Nai}}}, \ and\ \bibinfo
  {author} {\bibfnamefont {J.~T.~L.}\ \bibnamefont {Thong}},\ }\href@noop {}
  {\bibfield  {journal} {\bibinfo  {journal} {Nano Lett.}\ }\textbf {\bibinfo
  {volume} {14}},\ \bibinfo {pages} {3480} (\bibinfo {year}
  {2014})}\BibitemShut {NoStop}%
\bibitem [{\citenamefont {Foley}\ \emph {et~al.}(2015)\citenamefont {Foley},
  \citenamefont {Hern\'{a}ndez}, \citenamefont {Duda}, \citenamefont
  {Robinson}, \citenamefont {Walton},\ and\ \citenamefont
  {Hopkins}}]{foley2015}%
  \BibitemOpen
  \bibfield  {author} {\bibinfo {author} {\bibfnamefont {B.~M.}\ \bibnamefont
  {Foley}}, \bibinfo {author} {\bibfnamefont {S.~C.}\ \bibnamefont
  {Hern\'{a}ndez}}, \bibinfo {author} {\bibfnamefont {J.~C.}\ \bibnamefont
  {Duda}}, \bibinfo {author} {\bibfnamefont {J.~T.}\ \bibnamefont {Robinson}},
  \bibinfo {author} {\bibfnamefont {S.~G.}\ \bibnamefont {Walton}}, \ and\
  \bibinfo {author} {\bibfnamefont {P.~E.}\ \bibnamefont {Hopkins}},\ }\href
  {\doibase 10.1021/acs.nanolett.5b00381} {\bibfield  {journal} {\bibinfo
  {journal} {Nano Lett.}\ }\textbf {\bibinfo {volume} {15}},\ \bibinfo {pages}
  {4876} (\bibinfo {year} {2015})}\BibitemShut {NoStop}%
\bibitem [{\citenamefont {Sundaram}\ \emph {et~al.}(2011)\citenamefont
  {Sundaram}, \citenamefont {Steiner}, \citenamefont {Chiu}, \citenamefont
  {Engel}, \citenamefont {Bol}, \citenamefont {Krupke}, \citenamefont
  {Burghard}, \citenamefont {Kern},\ and\ \citenamefont
  {Avouris}}]{Sundaram2011}%
  \BibitemOpen
  \bibfield  {author} {\bibinfo {author} {\bibfnamefont {R.~S.}\ \bibnamefont
  {Sundaram}}, \bibinfo {author} {\bibfnamefont {M.}~\bibnamefont {Steiner}},
  \bibinfo {author} {\bibfnamefont {H.-Y.}\ \bibnamefont {Chiu}}, \bibinfo
  {author} {\bibfnamefont {M.}~\bibnamefont {Engel}}, \bibinfo {author}
  {\bibfnamefont {A.~A.}\ \bibnamefont {Bol}}, \bibinfo {author} {\bibfnamefont
  {R.}~\bibnamefont {Krupke}}, \bibinfo {author} {\bibfnamefont
  {M.}~\bibnamefont {Burghard}}, \bibinfo {author} {\bibfnamefont
  {K.}~\bibnamefont {Kern}}, \ and\ \bibinfo {author} {\bibfnamefont
  {P.}~\bibnamefont {Avouris}},\ }\href {\doibase 10.1021/nl201907u} {\bibfield
   {journal} {\bibinfo  {journal} {Nano Lett.}\ }\textbf {\bibinfo {volume}
  {11}},\ \bibinfo {pages} {3833} (\bibinfo {year} {2011})}\BibitemShut
  {NoStop}%
\bibitem [{\citenamefont {Ifuku}\ \emph {et~al.}(2013)\citenamefont {Ifuku},
  \citenamefont {Nagashio}, \citenamefont {Nishimura},\ and\ \citenamefont
  {Toriumi}}]{Ifuku2013}%
  \BibitemOpen
  \bibfield  {author} {\bibinfo {author} {\bibfnamefont {R.}~\bibnamefont
  {Ifuku}}, \bibinfo {author} {\bibfnamefont {K.}~\bibnamefont {Nagashio}},
  \bibinfo {author} {\bibfnamefont {T.}~\bibnamefont {Nishimura}}, \ and\
  \bibinfo {author} {\bibfnamefont {A.}~\bibnamefont {Toriumi}},\ }\href
  {\doibase 10.1063/1.4815990} {\bibfield  {journal} {\bibinfo  {journal}
  {Appl. Phys. Lett.}\ }\textbf {\bibinfo {volume} {103}},\ \bibinfo {pages}
  {033514} (\bibinfo {year} {2013})}\BibitemShut {NoStop}%
\bibitem [{\citenamefont {Barraza-Lopez}\ \emph {et~al.}(2010)\citenamefont
  {Barraza-Lopez}, \citenamefont {Vanevi\'{c}}, \citenamefont {Kindermann},\
  and\ \citenamefont {Chou}}]{Barraza2010}%
  \BibitemOpen
  \bibfield  {author} {\bibinfo {author} {\bibfnamefont {S.}~\bibnamefont
  {Barraza-Lopez}}, \bibinfo {author} {\bibfnamefont {M.}~\bibnamefont
  {Vanevi\'{c}}}, \bibinfo {author} {\bibfnamefont {M.}~\bibnamefont
  {Kindermann}}, \ and\ \bibinfo {author} {\bibfnamefont {M.~Y.}\ \bibnamefont
  {Chou}},\ }\href {\doibase 10.1103/PhysRevLett.104.076807} {\bibfield
  {journal} {\bibinfo  {journal} {Phys. Rev. Lett.}\ }\textbf {\bibinfo
  {volume} {104}},\ \bibinfo {pages} {076807} (\bibinfo {year}
  {2010})}\BibitemShut {NoStop}%
\bibitem [{\citenamefont {Liu}(2013)}]{liu2013}%
  \BibitemOpen
  \bibfield  {author} {\bibinfo {author} {\bibfnamefont {M.-H.}\ \bibnamefont
  {Liu}},\ }\href {\doibase 10.1007/s10825-013-0456-9} {\bibfield  {journal}
  {\bibinfo  {journal} {J. Comput. Electron.}\ }\textbf {\bibinfo {volume}
  {12}},\ \bibinfo {pages} {188} (\bibinfo {year} {2013})}\BibitemShut
  {NoStop}%
\bibitem [{\citenamefont {Lee}\ \emph {et~al.}(2008)\citenamefont {Lee},
  \citenamefont {Balasubramanian}, \citenamefont {Weitz}, \citenamefont
  {Burghard},\ and\ \citenamefont {Kern}}]{Kern2008}%
  \BibitemOpen
  \bibfield  {author} {\bibinfo {author} {\bibfnamefont {E.~J.~H.}\
  \bibnamefont {Lee}}, \bibinfo {author} {\bibfnamefont {K.}~\bibnamefont
  {Balasubramanian}}, \bibinfo {author} {\bibfnamefont {R.~T.}\ \bibnamefont
  {Weitz}}, \bibinfo {author} {\bibfnamefont {M.}~\bibnamefont {Burghard}}, \
  and\ \bibinfo {author} {\bibfnamefont {K.}~\bibnamefont {Kern}},\ }\href
  {\doibase 10.1038/nnano.2008.172} {\bibfield  {journal} {\bibinfo  {journal}
  {Nat. Nanotechnol.}\ }\textbf {\bibinfo {volume} {3}},\ \bibinfo {pages}
  {486} (\bibinfo {year} {2008})}\BibitemShut {NoStop}%
\bibitem [{\citenamefont {Mueller}\ \emph {et~al.}(2009)\citenamefont
  {Mueller}, \citenamefont {Xia}, \citenamefont {Freitag}, \citenamefont
  {Tsang},\ and\ \citenamefont {Avouris}}]{Mueller2010}%
  \BibitemOpen
  \bibfield  {author} {\bibinfo {author} {\bibfnamefont {T.}~\bibnamefont
  {Mueller}}, \bibinfo {author} {\bibfnamefont {F.}~\bibnamefont {Xia}},
  \bibinfo {author} {\bibfnamefont {M.}~\bibnamefont {Freitag}}, \bibinfo
  {author} {\bibfnamefont {J.}~\bibnamefont {Tsang}}, \ and\ \bibinfo {author}
  {\bibfnamefont {P.}~\bibnamefont {Avouris}},\ }\href {\doibase
  10.1103/PhysRevB.79.245430} {\bibfield  {journal} {\bibinfo  {journal} {Phys.
  Rev. B}\ }\textbf {\bibinfo {volume} {79}},\ \bibinfo {pages} {245430}
  (\bibinfo {year} {2009})}\BibitemShut {NoStop}%
\bibitem [{\citenamefont {Yoon}\ \emph {et~al.}(2014)\citenamefont {Yoon},
  \citenamefont {Forsythe}, \citenamefont {Wang}, \citenamefont {Tombros},
  \citenamefont {Watanabe}, \citenamefont {Taniguchi}, \citenamefont {Hone},
  \citenamefont {Kim},\ and\ \citenamefont
  {Ham}}]{kinetic_inductance_philipkim}%
  \BibitemOpen
  \bibfield  {author} {\bibinfo {author} {\bibfnamefont {H.}~\bibnamefont
  {Yoon}}, \bibinfo {author} {\bibfnamefont {C.}~\bibnamefont {Forsythe}},
  \bibinfo {author} {\bibfnamefont {L.}~\bibnamefont {Wang}}, \bibinfo {author}
  {\bibfnamefont {N.}~\bibnamefont {Tombros}}, \bibinfo {author} {\bibfnamefont
  {K.}~\bibnamefont {Watanabe}}, \bibinfo {author} {\bibfnamefont
  {T.}~\bibnamefont {Taniguchi}}, \bibinfo {author} {\bibfnamefont
  {J.}~\bibnamefont {Hone}}, \bibinfo {author} {\bibfnamefont {P.}~\bibnamefont
  {Kim}}, \ and\ \bibinfo {author} {\bibfnamefont {D.}~\bibnamefont {Ham}},\
  }\href {\doibase 10.1038/nnano.2014.112} {\bibfield  {journal} {\bibinfo
  {journal} {Nat. Nanotechnol.}\ }\textbf {\bibinfo {volume} {9}},\ \bibinfo
  {pages} {594} (\bibinfo {year} {2014})}\BibitemShut {NoStop}%
\bibitem [{\citenamefont {Nair}\ \emph {et~al.}(2008)\citenamefont {Nair},
  \citenamefont {Blake}, \citenamefont {Grigorenko}, \citenamefont {Novoselov},
  \citenamefont {Booth}, \citenamefont {Stauber}, \citenamefont {Peres},\ and\
  \citenamefont {Geim}}]{nair2008}%
  \BibitemOpen
  \bibfield  {author} {\bibinfo {author} {\bibfnamefont {R.~R.}\ \bibnamefont
  {Nair}}, \bibinfo {author} {\bibfnamefont {P.}~\bibnamefont {Blake}},
  \bibinfo {author} {\bibfnamefont {A.~N.}\ \bibnamefont {Grigorenko}},
  \bibinfo {author} {\bibfnamefont {K.~S.}\ \bibnamefont {Novoselov}}, \bibinfo
  {author} {\bibfnamefont {T.~J.}\ \bibnamefont {Booth}}, \bibinfo {author}
  {\bibfnamefont {T.}~\bibnamefont {Stauber}}, \bibinfo {author} {\bibfnamefont
  {N.~M.~R.}\ \bibnamefont {Peres}}, \ and\ \bibinfo {author} {\bibfnamefont
  {A.~K.}\ \bibnamefont {Geim}},\ }\href {\doibase 10.1126/science.1156965}
  {\bibfield  {journal} {\bibinfo  {journal} {Science}\ }\textbf {\bibinfo
  {volume} {320}},\ \bibinfo {pages} {1308} (\bibinfo {year}
  {2008})}\BibitemShut {NoStop}%
\bibitem [{\citenamefont {Datta}(1995)}]{dattabook}%
  \BibitemOpen
  \bibfield  {author} {\bibinfo {author} {\bibfnamefont {S.}~\bibnamefont
  {Datta}},\ }\href@noop {} {\emph {\bibinfo {title} {{Electronic transport in
  mesoscopic systems}}}}\ (\bibinfo  {publisher} {Cambridge University Press},\
  \bibinfo {year} {1995})\BibitemShut {NoStop}%
\bibitem [{\citenamefont {Nazarov}\ and\ \citenamefont
  {Blanter}(2009)}]{nazarov}%
  \BibitemOpen
  \bibfield  {author} {\bibinfo {author} {\bibfnamefont {Y.~V.}\ \bibnamefont
  {Nazarov}}\ and\ \bibinfo {author} {\bibfnamefont {Y.~M.}\ \bibnamefont
  {Blanter}},\ }\href@noop {} {\emph {\bibinfo {title} {Quantum Transp. Introd.
  to Nanosci.}}}\ (\bibinfo  {publisher} {Cambridge University Press},\
  \bibinfo {year} {2009})\BibitemShut {NoStop}%
\end{thebibliography}%
\end{document}